\documentclass[aps,pre,twocolumn]{revtex4}
\usepackage{graphicx} 
\usepackage{tikz-cd}
\usepackage{amsmath}
\usepackage{amsfonts}
\usepackage{amssymb}
\usepackage{adjustbox}
\usepackage{appendix}
\usepackage{xcolor}

\begin{document}
\title{Geometric Frustration in Twist-Bend Nematic Droplets}
\author{Joseph Pollard}
\email{joe.pollard@unsw.edu.au}
\affiliation{School of Physics, UNSW, Sydney, NSW 2052, Australia.}
\affiliation{EMBL Australia Node in Single Molecule Science, School of Medical Sciences, UNSW, Sydney, NSW 2052, Australia.}
\author{Richard G. Morris}
\email{r.g.morris@unsw.edu.au}
\affiliation{School of Physics, UNSW, Sydney, NSW 2052, Australia.}
\affiliation{EMBL Australia Node in Single Molecule Science, School of Medical Sciences, UNSW, Sydney, NSW 2052, Australia.}
\affiliation{ARC Centre of Excellence for the Mathematical Analysis of Cellular Systems, UNSW Node, Sydney, NSW 2052, Australia.}

\begin{abstract}
    Liquid crystals formed of bent-core molecules are exotic materials that exhibit the twist-bend nematic phase. This arises when an energetic preference for nonzero local bend distortion is accommodated via twist in the texture, resulting in properties synonymous with both smectics and cholesterics. Here we describe how the frustration inherent to the twist-bend phase can be exacerbated by confinement and boundary anchoring. Using a combination of numerical simulations, topological and geometric analysis, we catalogue the equilibrium textures that arise in spherical twist-bend droplets with a radial anchoring as the two key parameters---the molecular cone angle and the ratio between the pitch length and droplet radius---are varied. This form of confinement is known to produce a wide variety of topologically and geometrically complex metastable states in cholesterics. We find that twist-bend nematic droplets are no different, exhibiting a variety of complex layered states, defect constellations, and Hopfions. However, whilst many of the structures and defect configurations that we observe are equivalent to their cholesteric counterparts, they are geometrically very distinct, in part due to the absence of chirality. 
\end{abstract}

\maketitle

The twist-bend nematic is a liquid crystal mesophase in which the nematic orientation exhibits a heliconical modulation~\cite{dozov_spontaneous_2001, shamid_statistical_2013, jakli_physics_2018, binysh_geometry_2020, jakli_defects_2023}. It occurs in materials where the constituent molecules have a bent-core or `banana' shape, and is characterized by a state of nonzero bend distortion. On scales that are large compared to the helical pitch the twist-bend phase has the same elastic energy as a smectic and exhibits all the same features, textures, and defects. Conversely, for geometric reasons maintaining a nonzero bend distortion also requires twist and splay distortions~\cite{binysh_geometry_2020, pollard_intrinsic_2021, da_silva_moving_2021, selinger_director_2022}, with the former favouring the generation of structures usually associated with chiral liquid crystals, also called cholesterics. Accordingly, structures in twist-bend nematics display characteristics associated with both smectics and cholesterics~\cite{binysh_geometry_2020}. Notably, despite the constituent molecules of a twist-bend nematic being achiral, they exhibit stabile chiral structures such as Skyrmion lattices and even blue phases~\cite{subert_achiral_2024}. 

Like cholesterics, twist-bend nematics are inherently in a state of geometric frustration due to the impossibility of pure bend distortions~\cite{sadoc_liquid_2020, pollard_intrinsic_2021, da_silva_moving_2021, selinger_director_2022}. The inherent geometric frustration of cholesterics is partly responsible for the wide variety of metastable structures they display~\cite{sethna_relieving_1983, chen_generating_2013, tai_three-dimensional_2019, wu_hopfions_2022}, but they can also be subjected to confinement which further frustrates their ability to form textures with a constant sense of handedness, or `twist', with fascinating results. A natural experimental setting consists of a spherical droplet or shell of cholesteric material, with either planar~\cite{sec_geometrical_2012, darmon_topological_2016, darmon_waltzing_2016} or radial~\cite{posnjak_points_2016, posnjak_hidden_2017, posnjak_topological_2018, pollard_point_2019} anchoring at the boundary. The radial boundary condition is especially frustrating as it induces a topological constraint that prevents the material from twisting with a consistent sense of handedness throughout the droplet~\cite{pollard_point_2019}. By forcing regions of `wrong' handedness in a neighbourhood of the boundary, the anchoring `pins' hedgehog defects in the boundary region and leads to the stability of high-charge defects in the interior~\cite{pollard_point_2019, pollard_escape_2024}, among many other interesting textures. Thus, topological and geometric frustration produces a wide array of fascinating structures in chiral materials. 

The goal of this paper is to study the stable structures that arise in twist-bend nematic droplets with radial anchoring, and to compare them with the stable structures of cholesteric droplets, with a particular focus on the role played by the geometry of bend. Twist-bend nematics adopt twisted structures as a result of favouring helical directors, but as they are not chiral there is no distinction between left- and right-handed helices, with both having the same energy. The radial boundary condition on the droplet exacerbates the inherent frustration in twist-bend nematics because it is associated with splay, forcing both the bend and twist distortions to vanish close to the boundary. It is therefore reasonable to expect that radially-anchored droplet confinement will lead to large numbers of metastable textures in a twist-bend nematic, as it does for a cholesteric. 

To study the stable structures than can arise in droplets of twist-bend nematics, we adopt two approaches. Firstly, we determine the equilibrium structures that arise when the material is initialised with particular initial conditions that are (close to) stable equilibrium structures in cholesterics, covering all of the different point defect and soliton configurations as well as the different layer geometries observed in cholesterics. In each case we (i) determine whether the structure can be stabilised in some parameter regime in simulations of twist-bend nematics, and (ii) compare the resulting equilibrium state with its cholesteric equivalent. Secondly, we perform simulated quenches down from an isotropic phase into the twist-bend nematic in order to determine which of the equilibrium structures is likely to be observed in an experiment. In both cases we explore a wide parameter regime, attempting to cover all physically reasonable scenarios.

\section{Modelling of Twist-Bend Nematics and Cholesterics}
\begin{figure}[t]
\centering
\includegraphics[width=0.35\textwidth]{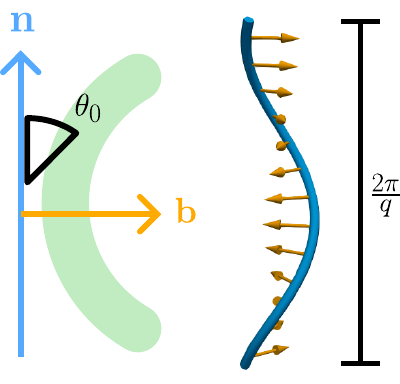}
\caption{Schematic diagram of a bent-core molecule (green) along with the desired director field ${\bf n}$ (blue) and bend direction ${\bf b}$ of the macroscopic phase. The cone angle $\theta_0$ is as indicated---a banana-shaped particle with a large molecular angle leads to a small cone angle with $\theta_0 \ll \pi/4$. Macroscopically, these phases organise into helices (blue curve) which undergo one full rotation over a pitch length $p$. This is related to the material's inverse pitch $q$ by $p = 2\pi/q$. The cone angle and pitch length are related to the elastic constants of our model by Eq.~\eqref{eq:parameters}. }
\label{fig0}
\end{figure}

Various approaches have been proposed for modelling materials that favour a nonzero value of the director distortions~\cite{kamien_liquids_1996, dozov_spontaneous_2001, shamid_statistical_2013, kats_landau_2014, pajak_nematic_2018, jakli_physics_2018}. We adopt the approach suggested by Meyer and developed by the Kent State group~\cite{shamid_statistical_2013} by introducing an auxiliary `polarisation field' ${\bf p}$ and coupling this to the desired director distortion. Microscopically, this polarisation field captures the `bent core' nature of the molecules as visualised in Fig.~\ref{fig0}, and macroscopically it sets a preferred nonzero value for the bend distortion.

The free energy to be minimised then involves terms penalising distortions in both the director field ${\bf n}$ and associated polarisation direction ${\bf p}$ with associated elastic constants $K, C$, as well as a term coupling polarisation to the bend distortion ${\bf b}$ of the nematic director and a constraint on the length of the polarisation, 
\begin{equation}\label{eq:tb_energy}
    E = \int \frac{K}{2}|\nabla {\bf n}|^2 + \frac{C}{2} |\nabla {\bf p}|^2 - \lambda {\bf p}\cdot {\bf b} + \frac{U}{4}|1-|{\bf p}|^2|^2.
\end{equation}
Note that our use of a director field for modelling the nematic order, rather than a Q-tensor, precludes the appearance of disclination lines. We will focus on point defect and soliton structures in this work, and leave disclination lines to a future study. 

Real materials exhibiting a twist-bend nematic phase have a strong anisotropy among the elastic constants, with the twist elastic constant being notably smaller than the others. The one constant approximation in Eq.~\eqref{eq:tb_energy} is therefore not really valid for these materials. However, we note that the cholesteric material employed in Refs.~\cite{posnjak_points_2016, posnjak_hidden_2017, posnjak_topological_2018} also had strong anisotropy among the elastic constants, but simulations performed with the one constant approximation still reproduce the observed structures, at least qualitatively~\cite{pollard_point_2019}. We therefore choose to make this assumption in order to reduce the number of parameters to be explored, and focus exclusively on the role played by the ratios $\lambda/K$ and $C/K$. Future studies could explore the role of elastic anisotropy, or choose to focus on a particular material whose elastic constants are known, e.g. one of those surveyed in Refs.~\cite{jakli_physics_2018, jakli_defects_2023}.

In a bulk material there is a natural ground state for this energy, given by the heliconical director field ${\bf n}_\text{h}$ and corresponding polarisation ${\bf p}_\text{h}$ exactly equal to its bend,
\begin{equation}\label{eq:heliconical_state}
    \begin{aligned}
        {\bf n}_\text{h} &= \cos  \theta_0 \,{\bf e}_z + \sin \, \theta_0 (\cos  qz \, {\bf e}_x + \sin  qz \,{\bf e}_y), \\
        {\bf p}_\text{h} &= -\sin qz  \,  {\bf e}_x + \cos  qz \, {\bf e}_y.
    \end{aligned}
\end{equation}
By taking the parameter $U$ in the free energy to be large, effectively imposing that the polarisation is unit length, we can relate the parameter ratios $\lambda/K, C/K$ in the free energy to to the cone angle $\theta_0$ and inverse pitch length $q$ in the heliconical state:
\begin{equation}\label{eq:parameters}
    \begin{aligned}
        \cos 2\theta_0 &= 1 + \frac{2C}{K} - \sqrt{\frac{4C}{K}\left(1 + \frac{C}{K} \right)}, \\
        q &= \frac{2\lambda}{K}\text{cot} \, 2 \theta_0.
    \end{aligned}
\end{equation}
These are obtained by substituting the helical director and its gradients into the energy Eq.~\eqref{eq:tb_energy} and minimising with respect to $\theta_0, q$, see Ref.~\cite{binysh_geometry_2020}. In particular, we note that the optimal cone angle is constrained to lie in the region $[0, \pi/4)$, as larger values of the cone angle require negative $\lambda, C$. This energy therefore models only materials with an oblique, and not acute, cone angle---the latter are more likely to form a splay-bend rather than twist-bend phase, and require a separate treatment. These parameters can be inferred in experiments from an examination of the ground state of an unconfined sample of the material, and what they represent is more intuitive than the parameters $\lambda, C$. We in turn reduce these to the consideration of a pair of dimensionless parameters $N > 0$ and $M > 4$, which are defined by,
\begin{equation}
        q = \frac{2N\pi}{R}, \ \ \ \ \ \ \ \ \ \
        \theta_0 = \frac{\pi}{M}.
\end{equation}
Note that our definition of the pitch length is $2\pi/q$, that is, the distance over which the polarisation field rotates through one full $2\pi$ turn, which differs from some other studies in which it is taken to be the half-pitch, $\pi/q$.

The other parameter in the model is $U$, which sets the scale of the bulk ordering energy. We explored the results of varying this parameter for several different fixed values of the other parameters. For sensible values of $U \in [0.01, 1]$ we observed no real difference in the equilibrium structures obtained, and thus we chose to keep this parameter fixed for all the simulations presented in this work. 

We compare our simulated structures with those in cholesterics, obtained by minimising the Frank free energy with a chiral term,
\begin{equation}\label{eq:ch_energy}
    E_\text{ch} = \frac{K}{2}\int |\nabla {\bf n}|^2 + 2q_0 {\bf n}\cdot \nabla \times {\bf n} + q_0^2.
\end{equation}

\begin{figure*}[t]
\centering
\includegraphics[width=0.98\textwidth]{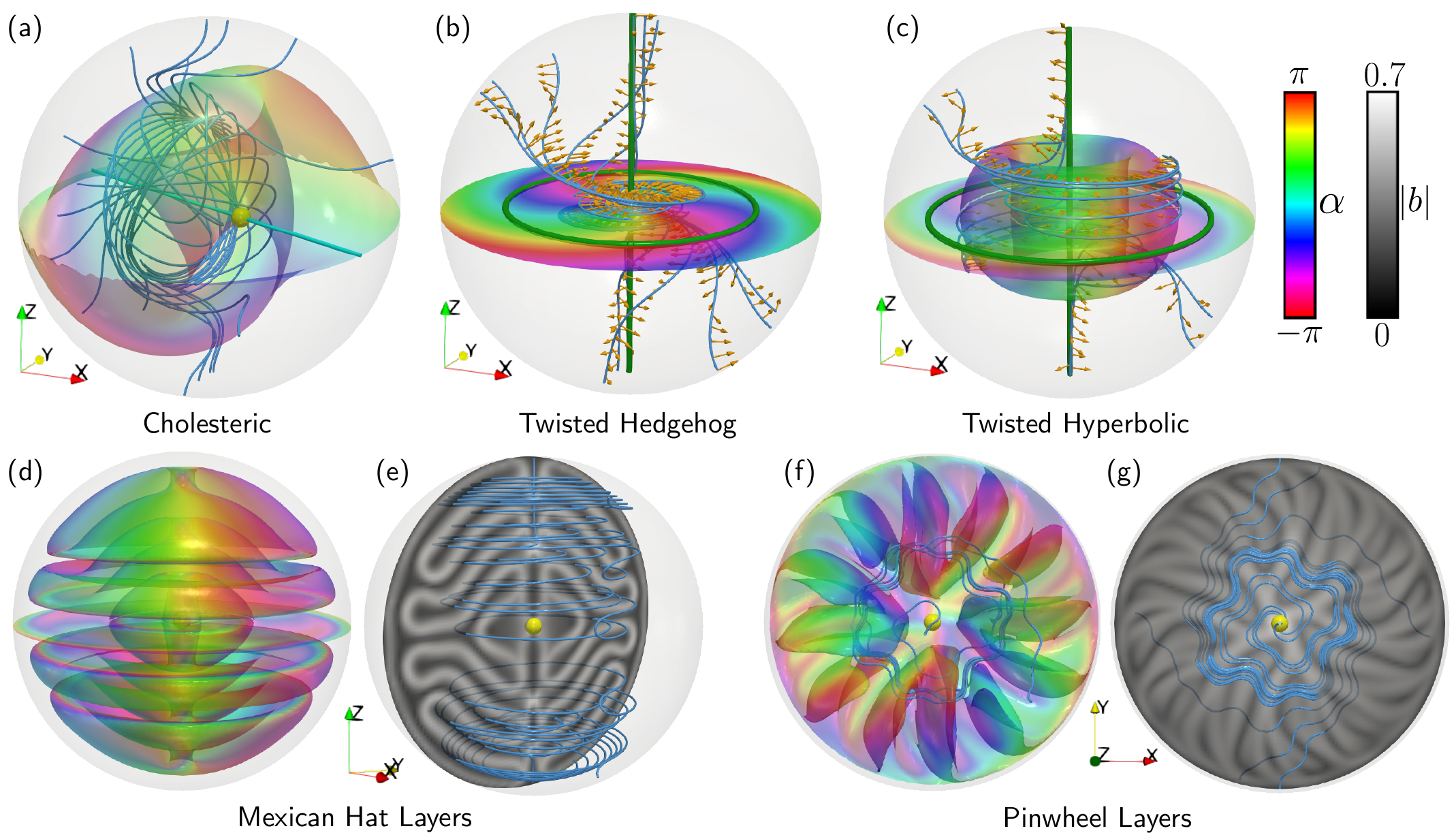}
\caption{Structures containing a central point defect. (a) In a cholesteric at $N=1$, a radial hedgehog (MI 0) undergoes a twist-driven conversion into a hyperbolic point defect (MI 2) that displaces slightly from the centre of a droplet. (b) In a twist bend nematic, at small $N$ (illustrated is $N=1, M=5$) the radial hedgehog begins twist slightly to introduce bend distortions. (c) At larger $N \geq 3$, (illustrated is $N=4, M=5$) the hedgehog (MI 0) converts to a hyperbolic defect (MI 2) in a bend-driven process. The defect remains at the centre of the droplet. Comparing with panel (a), we see the equilibrium texture is quite different from that of a cholesteric. (d,e) At large $N$ and small $M$ (illustrated in $N=12, M=5$) the droplet fills with pseudolayers in the shape of `Mexican hats'. The structure of the bend field and its zeros if very complex---we show the magnitude of the bend on a slice across the droplet. (f,g) At large $N$ and large $M$ (illustrated in $N=12, M=11$) the pseudolayers have a different geometry reminiscent of a `pinwheel'. The network of $\beta$-lines is again every complex. In each panel the defect position is indicated with a yellow sphere. The coloured surface is the PT surface with $n_z=0$, coloured according to the angle $\alpha = \tan^{-1}(n_y/n_x)$. Integral curves of the director are shown in blue, the bend direction in orange. In panel (a), the $\lambda$-line is indicated by a pale blue tube, while in panels (b,c) the $\beta$-lines are indicated by green tubes.}
\label{fig1}
\end{figure*}

Cholesterics have a single parameter, $q_0 = 2N \pi/R$, which sets the cholesteric inverse pitch length, and which is naturally analogous to the parameter $q$ for twist-bend nematics. In a droplet of radius $R$ we choose $q_0 = 2N\pi/R$ for various values of $N$---care should be taken when comparing with other studies, for example Refs.~\cite{posnjak_hidden_2017, posnjak_topological_2018, pollard_point_2019}, where the parameter $N$ is defined by the cholesteric half pitch, $q_0 = N\pi/R$, and therefore twice the value we use here. The energy minimizer is the cholesteric state,
\begin{equation}\label{eq:cholesteric_state}
    {\bf n}_\text{ch} = \cos  q_0 z \, {\bf e}_x + \sin  q_0z \,{\bf e}_y,
\end{equation}
which is of course the heliconical state in Eq.~\eqref{eq:heliconical_state} with $q=q_0$ and $\theta_0 = \pi/2$. 

We simulate structures starting from initial conditions which are close to equilibrium in cholesteric droplets---the exact initial conditions we use are detailed in the text. We also simulate quenches by initialising random director and polarisation fields. We explore parameters in the regime $N \in [1, 12]$ and $M \in [5, 12]$, covering a wide range of possible cone angles and pitch lengths---explicitly, increasing $N$ makes the pitch length shorter relative to the droplet radius, and increasing $M$ makes the cone angle smaller. Naively, small cone angles are expected to result in a more smectic-like texture, while larger cone angles are expected to result in a more cholesteric texture.

Typical pitch lengths of twist-bend phases are on the order of $10\text{nm}$, much shorter than the typical cholesteric pitch length of approximately $1\mu\text{m}$~\cite{jakli_physics_2018, jakli_defects_2023}. Our parameter regime therefore covers droplets of radius $10-120\text{nm}$. Converting to degrees, we cover cone angles in the range $15^\circ$ ($M=12$) to $36^\circ$ ($M=5$). As a representative example of a real material, Chen et. al.~\cite{chen_chiral_2013} report a twist-bend phase in the molecule CB(CH$_2$)$_7$CB with a pitch length of $8.3\text{nm}$ and a cone angle of about $24^\circ$ ($M=7.5$), comfortably within the parameter regime we explore. 

Additional details of our simulation methodology and our approach to visualising the equilibrium structures can be found in Appendix~\ref{appD}.

\section{Defect Configurations and Layered Structures in Twist-Bend Droplets}
\subsection{Structures with a Central Defect}
A droplet of achiral nematic with equal elastic constants and radial anchoring has a unique energy minimiser, given by a purely radial director field,
\begin{equation}
    {\bf n} = \frac{1}{r}(x{\bf e}_x + y{\bf e}_x + z{\bf e}_x).
\end{equation}
There is a hedgehog point defect at the centre of the droplet, and the director is twist and bend free, dominated by splay distortion. Orienting the director out from the boundary assigns this defect a charge of $+1$ and a Morse index (MI) of 0---see Refs.~\cite{pollard_point_2019, pollard_morse_2024} and Appendix~\ref{appA} for a definition of the MI. It is natural to begin our investigation with this structure. 

In chiral materials, the central MI 0 defect is unstable because it is impossible to maintain a consistent sense of handedness in a neighbourhood of the defect---the defect necessary sits on an interface between regions of positive and negative twist, or left and right handedness~\cite{pollard_point_2019}. At low values of the chirality the defect displaces towards the boundary to minimise the region where the sign of the twist is opposite to that which the material energetically prefers. At higher values, $N \geq 1$, the defect remains close to the centre of the droplet, but instead converts into a hyperbolic defect with MI 2 and a global structure resembling a `conch shell'~\cite{posnjak_hidden_2017, posnjak_topological_2018, pollard_point_2019}. This allows the texture to be uniformly chiral in the bulk, with a localised ring of reversed handedness near the boundary required by the boundary condition. The resulting structure in a cholesteric at $N=1$ is shown in Fig.~\ref{fig1}(a), and the conversion process is shown in Supplemental Movie 1.  

\subsubsection{Twisted Hedgehog Defect}
In a twist-bend nematic at $N=0$---in which case, the director and polarisation are decoupled---the radial hedgehog is stable and remains the equilibrium configuration for any value of the cone angle. First consider a large cone angle, $M = 5$. As we increase $N$ above zero the structure begins to change. While it remains a hedgehog with MI 0, the integral curves begin to bend, as for $N=1$ in Fig.~\ref{fig1}(b). In cholesterics, point defects in the director field must sit on defects in the pitch axis, called $\lambda$-lines. In twist-bend nematics, the equivalent line-like features are the $\beta$-lines, zeros of the bend direction---see Appendix~\ref{appC} and Ref.~\cite{binysh_geometry_2020} for a more detailed discussion of these features. In the structure shown in Fig.~\ref{fig1}(b), a single $\beta$-line with $+1$-winding---topologically required by the presence of the point defect---lies along the $z$-axis. An additional closed loop $\beta$-line, again with $+1$ winding, encloses this, and is associated with a helical perversion in the integral curves of the director field. The local structure of the director field close to these is that of a meron, although only the central line is actually the core of a meron tube, and there is no topological reason for the second $\beta$-line to be there: it is geometric in nature, a ring of inflection points in the integral curves of the director.

We refer to this structure as a `twisted hedgehog'. It is different from the displaced hedgehog texture observed in a cholesteric, but similar to structures observed in chromonic liquid crystals and other materials in which twist distortions are energetically cheaper than splay distortions~\cite{lavrentovich_phase_1986}. Structures similar to these were predicted to occur in twist-bend nematics by Kl\'eman~\cite{kleman_developable_1980}, and have been observed in experiments~\cite{nastishin_textural_2003}. 

At smaller cone angles ($M = 6-12$) the twisted hedgehog is the equilibrium defect configuration for the director for a wider range of $N$, specifically $N=2-6$. However, very small cone angles favour the production of additional $\beta$-lines. The need for these inflection points is a manifestation of the fact that a consistent sense of handedness (sign of the twist) cannot be achieved for radial anchoring---the integral curves of the director, which are helices, must change their handedness somewhere near the boundary. Overall, in all the textures studied in this work we observe many more $\beta$-lines in twist-bend materials than we do $\lambda$-lines in cholesterics, with most $\beta$-lines not being required by topological constraints.

\subsubsection{Twisted Hyperbolic Defect}

For $M=5$ and larger values of $N \geq 3$, the bending is strong enough to induce a Hopf bifurcation in the cross-section~\cite{ciuchi_inversion_2024, pollard_morse_2024}, and the defect undergoes a conversion from a hedgehog, MI 0, to a hyperbolic defect of the same charge, MI 2, as it would in a cholesteric. However, the equilibrium structure in a twist-bend nematic at $N=4$, Fig.~\ref{fig1}(c), is quite different from that in a cholesteric, Fig.~\ref{fig1}(a). The development of this structure is shown in Supplemental Movie 2.

\subsubsection{Mexican Hat Texture}

At $M=5$ and very large $N$ the structure changes substantially, generating nested families of pseudolayers with a complex structure. The pseudolayers nucleate along the central $\beta$-line and spread throughout the droplet, as shown in Supplemental Movie 3. We show the equilibrium texture for $N=12$ in Fig.~\ref{fig1}(d). The integral curves of the director field contain many limit cycles (closed loops), and trajectories either connect one of these loops to the boundary/defect, or else connect pairs of loops. A collection of integral curves are shown in Fig.~\ref{fig1}(e)---they are almost flat circles. 

In these cases, the pseudolayers---revealed through an examination of a Pontryagin--Thom (PT) surface, and also the magnitude of the bend distortion, as detailed in Appendix~\ref{appD}---resemble a series of nested `Mexican hats'. The structure of the bend field is accordingly very complex: we show its magnitude across a slice in Fig.~\ref{fig1}(e). The network of $\beta$-lines in these droplets is far more complex than the network of $\lambda$-lines in cholesteric droplets---for clarity of visualisation we do not show them. 

\begin{figure*}[t]
\centering
\includegraphics[width=0.98\textwidth]{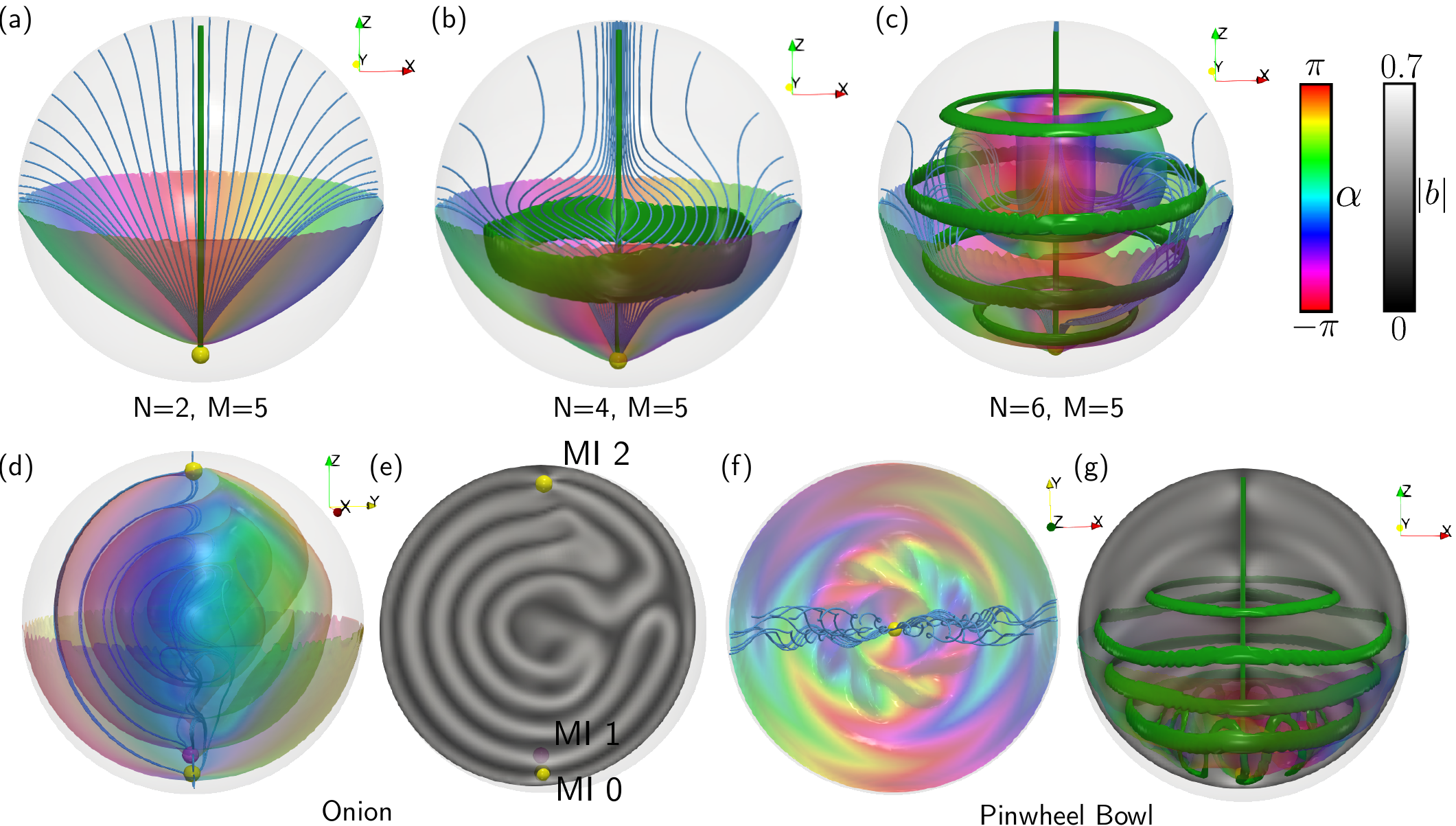}
\caption{Structures containing a boundary point defect. (a-c) At $M=5$ and small $N=2-6$, the structures resemble the twisted hedgehog. (d,e) Even at large $N$ (shown here is $N=10$) the boundary defect remains MI 0. However, the droplet fills with additional bowl-shaped and toroidal layers which resemble the petals of a flower. The integral curves of the director are folded arcs on these layers. (f,g) At $N=10, M=11$ we still observe a bowl structure, with with `ridges' in the PT surface analogous to the pinwheel. There are many closed-loop $\beta$-lines, one of which has a complex geometry which is also seen in the pinwheel structure. In each panel the defect position is indicated with a yellow sphere. The coloured surface is the PT surface with $n_z=0$, coloured according to the angle $\alpha= \tan^{-1}(n_y/n_x)$. Integral curves of the director are shown in blue. In panels (a-c) and (g) the $\beta$-lines are indicated by green tubes---we omit the $\beta$-lines in other panels for clarity. In panels (e,g) the magnitude of the bend is shown on a slice.}
\label{fig2}
\end{figure*}

\subsubsection{Pinwheel Texture}

At larger values of $N = 6-12$ and $$M = 6-12$$, the central defect again undergoes a conversion to MI 2 and the droplet again fills with pseudolayers, although these have a very different configuration to those that occur at small $M$. The layers form a `pinwheel structure', and the integral curves of the director are helices with multiple, less dramatic kinks than those observed at $M=5$. We show this for $N=12, M=11$ in Fig.~\ref{fig1}(f). The magnitude of the bend is shown on a slice in Fig.~\ref{fig1}(g). 

In these textures, and for the rest of the structures shown throughout, the structure of the polarisation field is similar to that of the bend. However, while the bend is constrained to be orthogonal to the director field the polarisation is not, and thus it is able to `escape' the singular $\beta$-lines of the bend and replace them with either nonsingular integral curves or strings of singularities. The plots of $|{\bf b}|$ broadly show the structure of the polarisation field in each case---we do not further describe the polarisation field in this work.

\subsection{Layered Structures with a Single Boundary Defect}
By far the most common structures in cholesteric droplets with radial anchoring are those that contain a single point defect close to the boundary~\cite{posnjak_topological_2018}. These can arise from a hedgehog displaced towards the boundary, which occurs at small $N$. An alternative is to have the droplet be filled with either flat layers (with a normal along the $z$-direction, say), cylindrical layers (with normal along the radial direction in a cylindrical coordinate system), or spherical layers---the latter case is known as the Frank--Pryce structure, and we will come to it presently. All three layer geometries can occur in cholesteric droplets with planar anchoring without a need for director defects in the bulk, and simply a pair of boojums on the boundary~\cite{sec_geometrical_2012}. In a droplet with radial anchoring, the layered structures are accompanied by a single hedgehog defect, which meets both the topological requirement to have a total $+1$ defect charge, as well as the additional topological requirement of a region of reversed handedness close to the boundary~\cite{pollard_point_2019}. Spherical layers with a string of defects on an axis through the droplet are also possible, and we will examine these in the next subsection.

\subsubsection{Boundary Hedgehogs, the Bowl and Onion Textures}

We can initialise a director field with a boundary hedgehog (MI 0 defect) by taking the director to be ${\bf n} = {\bf e}_z$ in the interior and imposing radial boundary conditions. In a cholesteric, the defect remains a hedgehog close to the boundary for small $N$, but for $N \geq 2$ the defect moves back into the interior and converts to a hyperbolic defect, as in Fig.~\ref{fig1}(a). 

In a twist-bend nematic at $M=5$ and small $N$ this results in a `bowl'-like structure with a $\beta$-line passing through the defect, Fig.~\ref{fig2}(a). As we increase $N$ with $N \leq 6$, Fig.~\ref{fig2}(b,c), the integral curves become more twisted up, and additional closed loop $\beta$-lines nucleate. These lines are inflection points in the director's integral curves resulting from the frustration of trying to pack helices into the droplet. Unlike for the case of a central point defect, the boundary defect remains a hedgehog and does not convert to a hyperbolic defect. Eventually we generate an additional, toroidal component to the PT surface, as in Fig.~\ref{fig2}(c). A toroidal part of the PT surface is often the signature of a Hopfion~\cite{chen_generating_2013, wu_hopfions_2022}, and a casual glance at the structure in Fig.~\ref{fig2}(c) may lead us to conclude we have found such a soliton. However, an examination of the preimages of ${\bf e}_x$ shows there is no linking---see Appendix~\ref{appB}---so the Hopf invariant is zero. Although this texture does not contain a Hopfions, these kinds of soliton do arise in our simulations, and we will examine them presently. 

At large values of $N$, say $N=12$, the director nucleates a series of nested pseudolayers which produce an `onion'-like structure, Fig.~\ref{fig2}(d,e). It also nucleates a pair of additional defects, with MI 1 and MI 2. The integral curves of the director field form a series of U-shaped loops that leave the boundary MI 0 defect, bend sharply back on themselves, and then limit on either the boundary or another defect. Similar structures have been observed in experiments~\cite{eremin_light-responsive_2018}, and can be seen in a number of other textures we discuss in this work. 

We can appreciate the prevalence of these U-shaped integral curves by another analogy with cholesterics. Cholesterics are inherently geometrically frustrated, as the pure twist distortion mode they would like to adopt is not possible in flat space. This distortion mode is achievable in several curved spaces: notably, it can be achieved in the Lie groups $SU(2)$ (equivalently, the 3-sphere $\mathbb{S}^3$), the Heisenberg group, and $SL(2, \mathbb{R})$~\cite{sethna_relieving_1983, sadoc_liquid_2020, pollard_intrinsic_2021, da_silva_moving_2021}. The first two of these idealised structures correspond respectively to cholesteric double-twist and single-twist---the third has not been properly studied. Cholesterics try to achieve one of these idealised structures, but can do so only locally in flat space. The textures resulting from attempts to achieve the idealised double twist in a flat space are blue phases, meron/Skyrmion lattices, and Hopfions~\cite{wu_hopfions_2022}, familiar and interesting textures in cholesterics. 

\begin{figure}[t]
\centering
\includegraphics[width=0.35\textwidth]{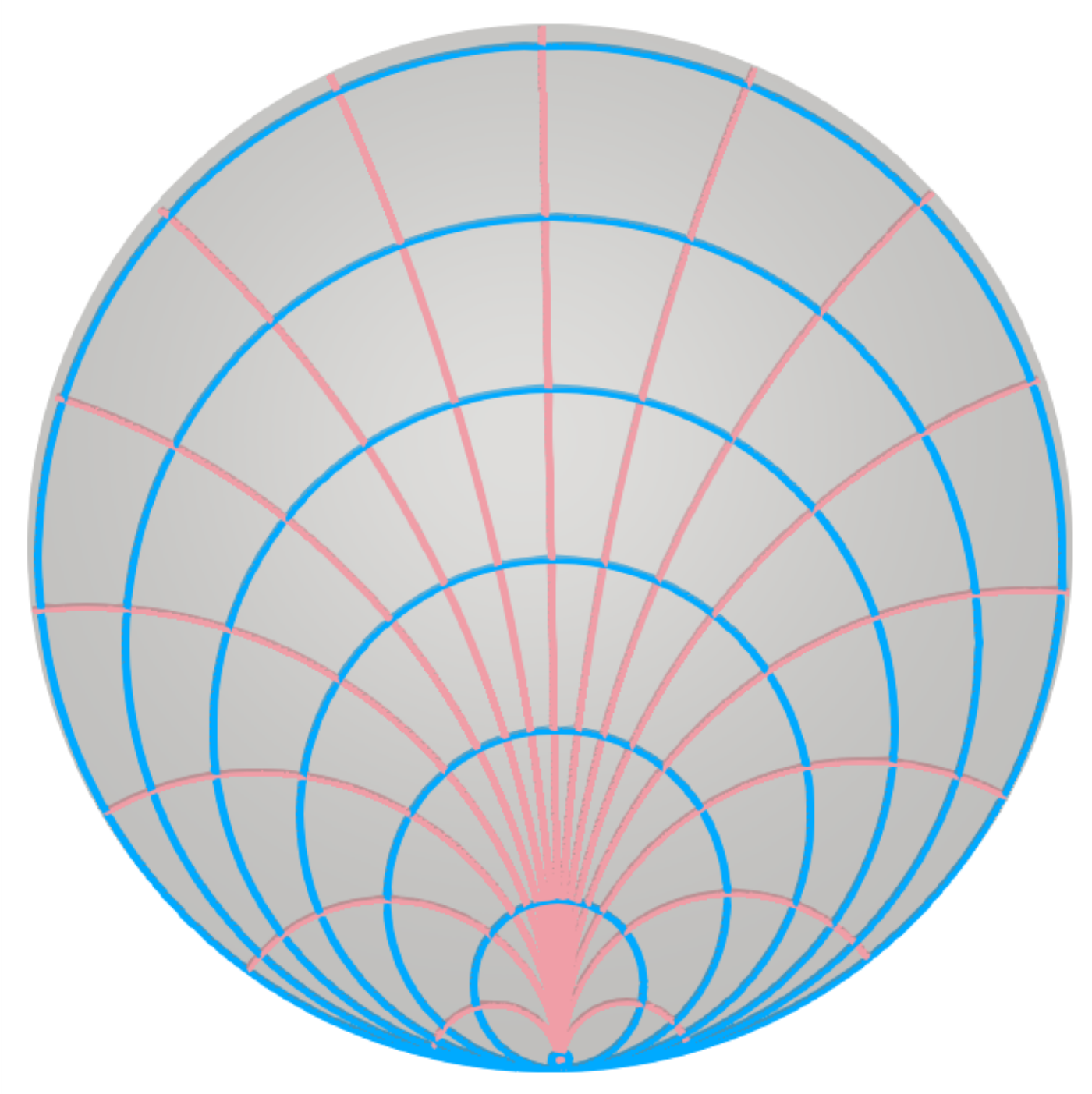}
\caption{The negatively curved hyperbolic space $\mathbb{H}^3$ admits director fields which have either pure constant splay or pure constant bend distortions; neither situation can be achieved in a flat space. There is a natural projection from $\mathbb{H}^3$ into the unit sphere in $\mathbb{R}^3$, the Poincar\'e ball representation. We show a cross-sectional disk of the Poincar\'e ball, along with the integral curves of the director fields obtained by pushing forward the pure bend (blue) and pure splay (red) along the projection. As these are director fields in flat Euclidean space they are frustrated, and no longer pure bend/splay. They also contain a $+1$ point defect, which sits on the boundary of the sphere. They give reasonable approximations to textures that may occur in confined twist-bend nematics. }
\label{fig3}
\end{figure}

Bent-core materials are similarly frustrated, as one cannot achieve a pure bend distortion in flat space, but we may likewise describe curved spaces in which pure bend distortions can be achieved. Examples include the 3D hyperbolic space $\mathbb{H}^3$, the cross product $\mathbb{H}^2 \times \mathbb{R}$ of the 2D hyperbolic space with a line, and the space $\text{Sol}$, which is the Type V Lie group in the Bianchi classification~\cite{sadoc_liquid_2020, pollard_intrinsic_2021, da_silva_moving_2021}. The hyperbolic space $\mathbb{H}^3$ admits a very natural map into a unit ball inside $\mathbb{R}^3$, the Poincaré ball representation, and this ball can be identified with the interior of our droplet. In this representation the pure-bend director in $\mathbb{H}^3$ maps onto a director in the unit ball with integral curves that are `horocycles' on a series of nested `horospheres' emanating from a `defect' point at the north pole of the ball. A cross sectional disk cutting across the Poincar\'e ball, foliated by horocycles (blue curves), is shown in Fig.~\ref{fig3}. While this is not exactly the texture observed in our simulations, it does suggest a possible explanation for the prevalence of the loop-like integral curves seen in Fig.~\ref{fig2}(d,e). There is also a pure splay director field in $\mathbb{H}^3$, which has nonzero bend when projected down into the unit ball in $\mathbb{R}^3$. This projection is shown as red curves in Fig.~\ref{fig3}, and we see that it resembles the small $N$ texture with a boundary defect, Fig.~\ref{fig2}(a). Although informal, these observations strongly suggest it would be fruitful to perform a more detailed geometric analysis of twist-bend structures in terms of hyperbolic geometry, just as spherical geometry allows for an analysis of cholesterics~\cite{sethna_relieving_1983}.  

For larger $M=11$ we see bowl-like textures for all values of $N$ surveyed. At large $N$ the PT surface develops ridges akin to those in the pinwheel texture, Fig.~\ref{fig2}(f,g). The boundary defect is a hedgehog.

All of these structures are analogous to the structures described with a centre defect---`mirroring' the boundary defect structures with the defect placed in the centre of the droplet produces the equivalent texture with a central defect.

\subsubsection{Flat Layers, the Octopus and Screw Textures}
\begin{figure*}[t]
\centering
\includegraphics[width=0.98\textwidth]{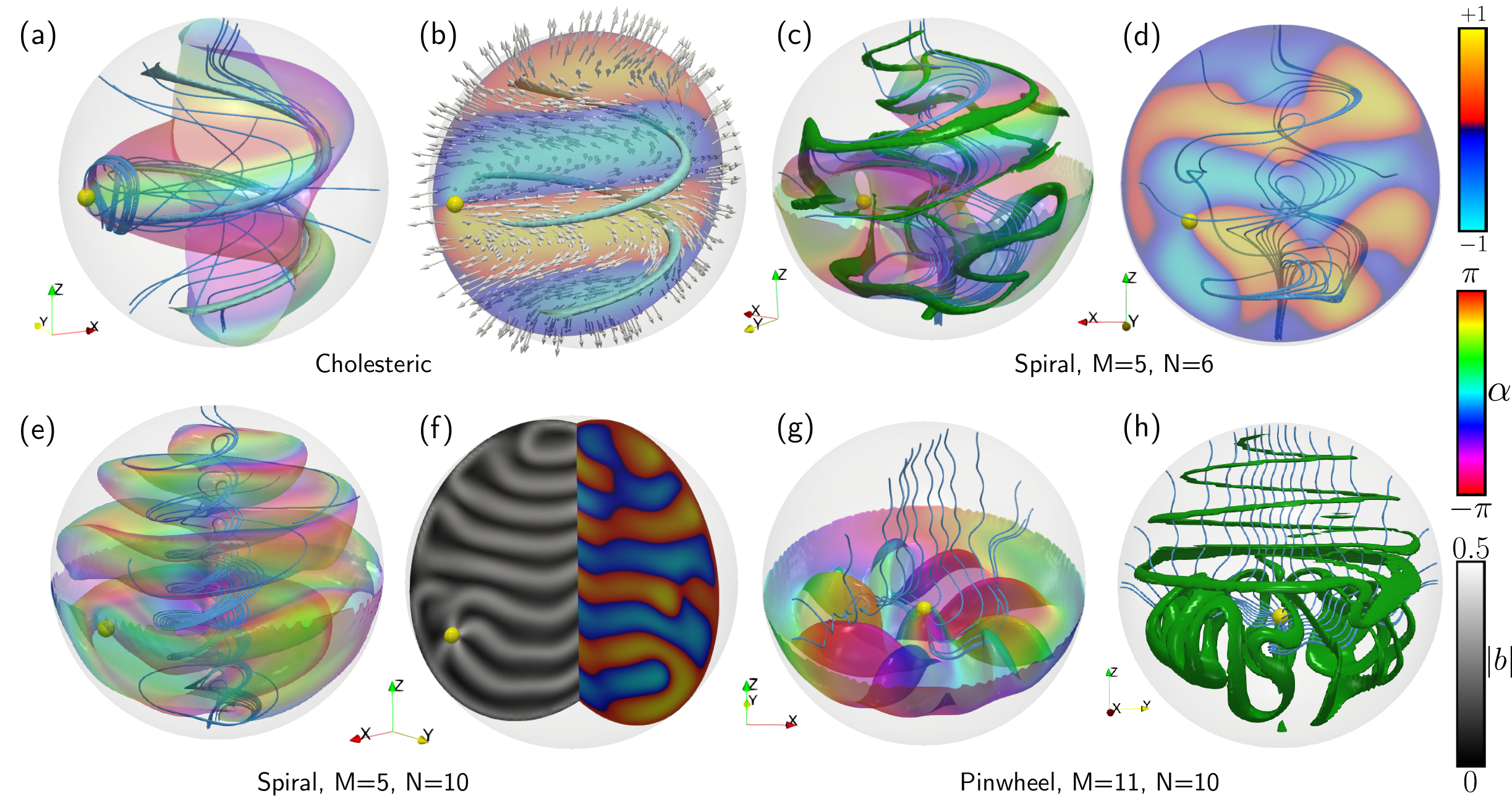}
\caption{Structures with flat layers. (a,b) A cholesteric at $N=1$. Flat layers aligned along the $z$ direction fill the interior of the droplet, with an additional boundary region containing a hedgehog defect. Two $\lambda$-lines (blue tubes) spiral out from the defect to the poles. In panel (b) we show the director field (white) and the dot product between the director and the normal to a slice. This clearly shows the layer structure. (c,d) Twist-bend nematics reproduce similar structures. At $M=5, N=6$, we see `spiral' structures similar to the flat cholesteric layers. The layer geometry, which is revealed by examining the dot product between the director and the normal to a slice (panel (d)), is different, but the topology is the same. The $\beta$-lines in the droplet spiral out from the defect. (e,f) For larger $N$, here $N=10$, the layered structure becomes more regular. (f) We can see the layers through an examination of the bend magnitude (black and white) or the dot product between the director and the normal to a slice  (blue and yellow). (g,h) At larger $M$, here $M=11$, spiral textures give way to variants on the pinwheel bowl structure, but with a different configuration to the $\beta$-lines, which retain the spiralling structure characteristic of flat layers. The coloured surfaces are the PT surfaces with $n_z=0$, coloured according to the angle $\alpha = \tan^{-1}(n_y/n_x)$. $\beta$-lines are shown as green tubes.}
\label{fig4}
\end{figure*}

We next consider flat layers, initialised by the director field
\begin{equation}
    {\bf n} = \cos(N\pi z/R){\bf e}_x + \sin(N\pi z/R){\bf e}_y.
\end{equation}
In a cholesteric, Fig.~\ref{fig4}(a,b), the layers (level sets of $z$) remain flat in the interior of the droplet. The defect (MI 0) appears on the $x$-axis, and two $\lambda$-lines spiral out from it and towards opposite poles on the sphere. Cholesteric $\lambda$-lines are dislocations in the layer structure, and these are required to match on to the boundary condition. 

In a twist-bend nematic at cone angle $M=5$ and small $N \leq 2$, an initial condition with flat layers equilibrates in a `bowl'-like structure with a single hedgehog point defect displaced from the centre towards the boundary. This is not a layered structure, rather it is the natural analogue of the cholesteric texture with a single displaced point defect. There are a pair of $\beta$-lines coiled up into helices, similar to the $\lambda$-lines in the cholesteric structure but in the bulk of the droplet, rather than trapped in a boundary layer. 

At we increase $N$, Fig.~\ref{fig4}(c,d), the PT surface becomes a screw-like structure similar to the cholesteric, and the layers become more apparent. The $\beta$-lines becoming more twisted up, similar to the $\lambda$-lines in the cholesteric. For very large $N \geq 10$, Fig.~\ref{fig4}(e,f), a clear `screw' emerges, with $\beta$-lines and integral curves of the director both forming helices, similar to the cholesteric structure. There is still only a single defect. We show the development of these textures in Supplemental Movies 4 and 5, covering $M=5, N=6$ and $M=5, N=12$ respectively.

At larger $M \geq 8$, we see only pinwheel bowl structures, similar to Fig.~\ref{fig2}(f,g). However, the configuration of $\beta$-lines in the droplet is slightly different: the multiple rings present in the texture in Fig.~\ref{fig2}(g) are merged into a single $\beta$-line loop that spiral around in a helix, replacing the axis running through the droplet. Indeed, this texture, shown in Figs.~\ref{fig4}(g,h), has only a single additional $\beta$-line beyond the ones that are topologically required by the point defect, a loop that with the complex undulating structure characteristic of the pinwheel-type textures. For intermediate values of the cone angle, $5 < M < 8$, we find textures that interpolate between the screw and the pinwheel bowl. 

\subsubsection{Cylindrical Layers, The Flower Texture}
\begin{figure*}[t]
\centering
\includegraphics[width=0.98\textwidth]{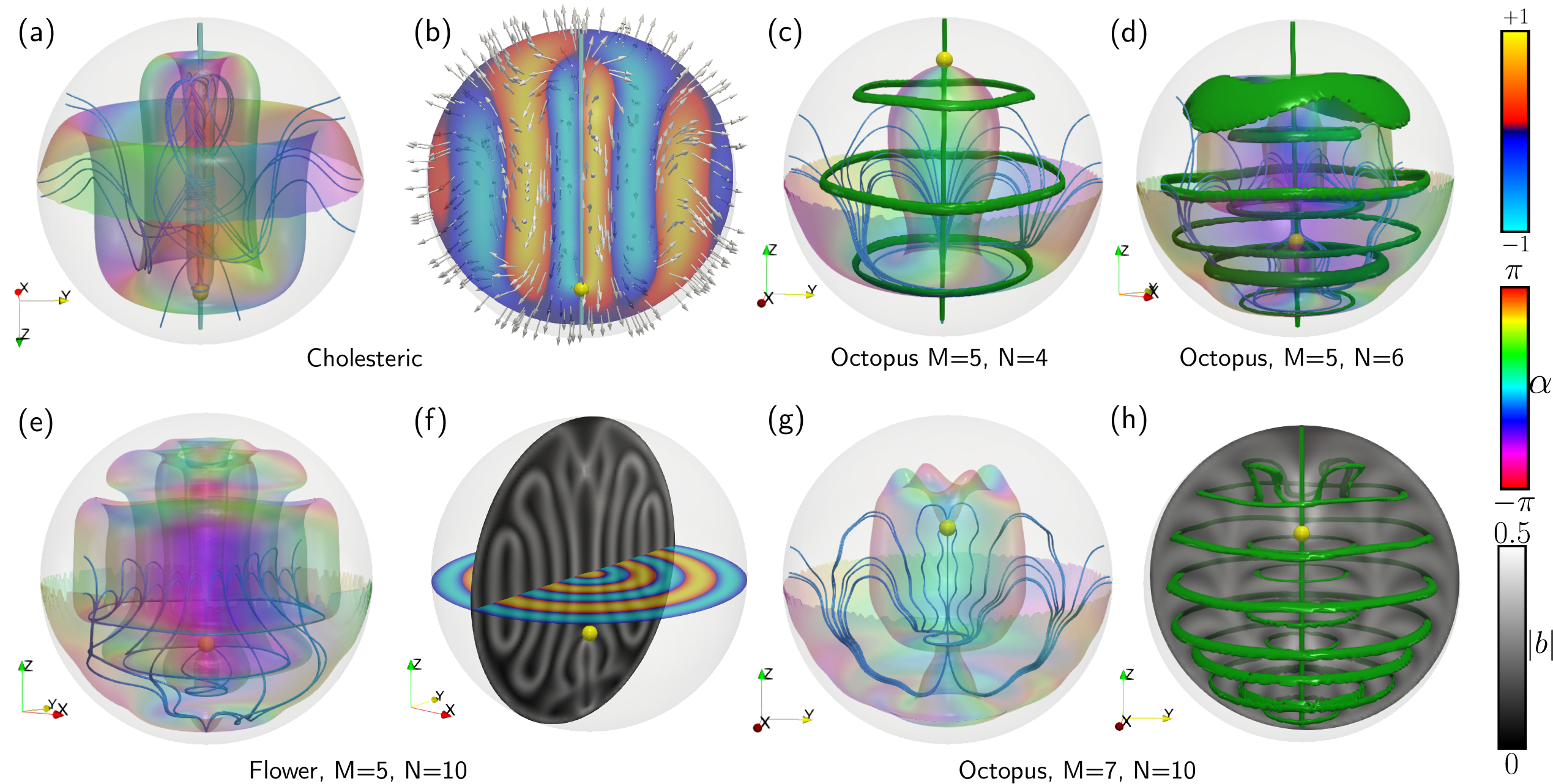}
\caption{Structures with cylindrical layers. (a,b) Cylindrical layers in a cholesteric droplet are the result of inserting a double-twist cylinder along an axis, here the $z$ axis. In a droplet with radial anchoring the texture must be accompanied by a hedgehog point defect, which is pinned to the boundary and sits on the axis of the double-twist cylinder, which is a $\lambda$-line (pale blue). We show a slice coloured according to the dot product between the director and slice normal (yellow and blue), with the director field shown as grey arrows. (c,d) Structures with approximately cylindrical layers occur in droplets of twist-bend nematic as well. At large cone angle and small $N$ we find `octopus' textures, which contain excess closed loop $\beta$ lines (green tubes) in addition to the topologically-required $\beta$-lines along the core axis. We show this structure at (c) $M=5, N=4$ and (d) $M=5, N=6$. (e,f) At larger $N$, here $N=5$, the equilibrium `flower' structures is visually much closer to the cholesteric analogue. The layers can be seen by looking at either the bend magnitude on a slice (black and white) or the dot product between the director and slice normal. (g,h) Flower structure do not occur from larger values of $M$, instead we see variants on the octopus, here at $M=7, N=10$. In each panel the coloured surface is the PT surface with $n_z=0$, coloured according to the angle $\alpha = \tan^{-1}(n_y/n_x)$. Integral curves of the director are shown in blue.}
\label{fig5}
\end{figure*}

Cylindrical layers are initialised using the director
\begin{equation}
    {\bf n} = -\cos(2N\pi r/R){\bf e}_z + \sin(2N\pi r/R){\bf e}_\rho,
\end{equation}
where $\rho = \sqrt{x^2+y^2}$. This initial condition is `escape down' along the $z$-axis, necessary to produce a right-handed texture~\cite{pollard_escape_2024}, but in a twist-bend material one could also consider `escape up' for a left handed texture. The cholesteric realisation of this texture at $N=4$ is shown in Fig.~\ref{fig5}(a,b). There is a single $\lambda$-line along the axis containing the point defect. 

In twist-bend nematics at small values of $N$ the equilibrium structures, regardless of cone angle, have a single hedgehog defect (MI 0) close to the boundary, with a PT surface resembling a `bowl', as in Fig.~\ref{fig2}(a). For large $M$ these are only types of structure we see at any $N \leq 12$ surveyed. For smaller $M=5$, increasing $N$ leads to the formation of octopus-like structures similar to what is shown in Fig.~\ref{fig2}(c)---see Fig.~\ref{fig5}(c,d) for $N=4$ and $N=6$. As we increase $N$ still further these textures begin to resemble complex `flowers', with cylindrical pseudolayers filling the droplet. The sole boundary defect is still a hedgehog. Fig.~\ref{fig5}(e,f) shows such a structure at $N=10$. The `petals' of the flower are layers for the director field, which has bending domains filling the spaces between. The $\beta$-lines consist of closed loops encircling a central axis. 

At intermediate cone angles, $M=7$, we see variants of the octopus structure, Fig.~\ref{fig5}(g,h). These structures exhibit the U-shaped integral curves and have large numbers of excess $\beta$-lines. 

We show the development of these textures at $M=5, N=6$, $M=5, N=12$, $M=11, N=6$, and $M=11, N=12$ respectively in Supplemental Movies 6-9. 

\subsubsection{Spherical Layers, the Frank--Pryce Texture}

An alternative structure in a cholesteric in a droplet with planar anchoring has spherical layers with a full line singularity of winding $+2$ connecting the boundary to the centre of the droplet. This line singularity splits into a pair of $+1$ singular lines, which then `escape' to form a pair of $\lambda$-lines and no director defects. These $\lambda$-lines must be wrapped helically around one another in order for the escape process to result in chiral texture~\cite{pollard_escape_2024}. For radial anchoring this structure may also occur, with an additional hedgehog defect at the boundary. This structure, known as the radial spherical structure (RSS) or Frank-Pryce texture~\cite{frank_multiplication_1950, bezic_structures_1992, sec_geometrical_2012, posnjak_topological_2018, pollard_escape_2024}, is well known in cholesterics and has possibly been observed in droplets of twist-bend nematic as well~\cite{eremin_light-responsive_2018, eremin_effects_2020, krishnamurthy_electric_2021, jakli_defects_2023}. 

The Frank--Pryce structure is initialised using the director field ${\bf n} = {\bf m}/|{\bf m}|$, where
\begin{equation}
    {\bf m} = \cos(N\pi r/2R) \, {\bf e}_1 + \sin(N\pi r /2R) \, {\bf e}_2,
\end{equation}
is written in terms of the two directions
\begin{equation}
    \begin{aligned}
        {\bf e}_1 &= 2\theta\phi {\bf e}_\theta + (\theta^2-\phi^2){\bf e}_\phi, \\
        {\bf e}_2 &= 2\theta\phi {\bf e}_\phi - (\theta^2-\phi^2){\bf e}_\theta.
    \end{aligned}
\end{equation}
We have chosen the argument $N\pi r/2R$ for the number of rotations because this selects for a number of layers ($N/2$) corresponding to the observed high $N$ equilibrium in the twist-bend material, as we describe momentarily. Initialising a different layer number leads to a reconfiguration that still results in $N/2$ layers at equilibrium.

\begin{figure*}[t]
\centering
\includegraphics[width=0.98\textwidth]{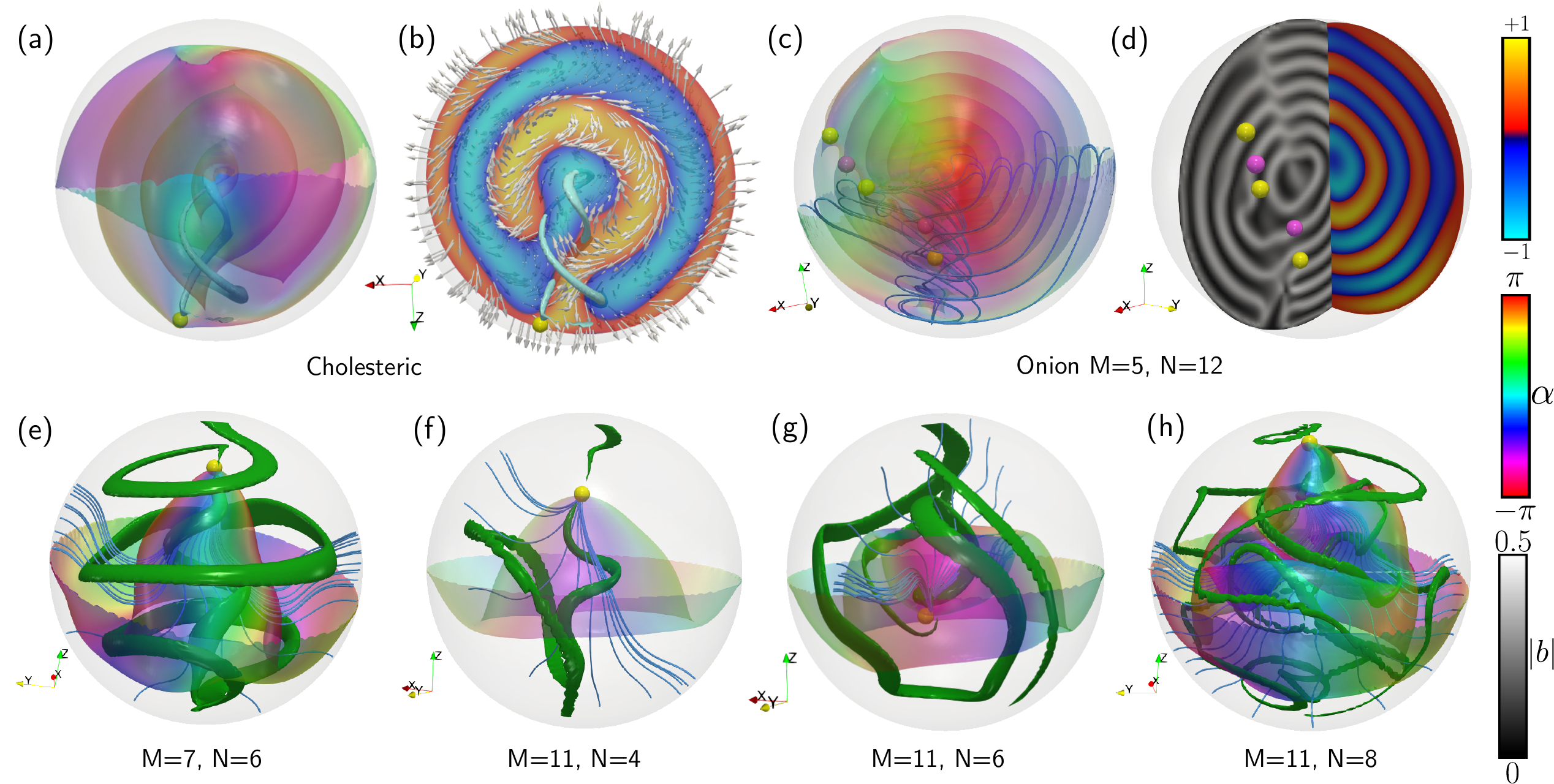}
\caption{The Frank--Pryce, or RSS, texture. (a,b) This texture is common in cholesteric droplets. Shown here is $N=4$. The characteristic features of this texture are the double helix of $\lambda$-lines (pale blue tubes) and the spherical layers, which are clearly seen in (b), where we plot the dot product between the director field and the normal to a slice along with the director field itself (grey arrows). (c,d) This structure is also stable in twist-bend droplets at large cone angles and large $N$---here we show $M=5, N=12$. The twist-bend structure nucleates additional defects, and here all defects are hyperbolic with either MI 1 (f$-1$ charge, purple) or MI 2 ($+1$ charge, yellow). (e-h) We did not observe structures with spherical layers from $M \geq 7$ at any value of $N$ surveyed. Instead, we observe structures with point defects in the bulk of the droplet and spiralling helices of $\beta$-lines (green tubes). Typically there are only two, those topologically required by the presence of the point defect. In each panel the coloured surface is the PT surface with $n_z=0$, coloured according to the angle $\alpha = \tan^{-1}(n_y/n_x)$. Integral curves of the director are shown in blue.}
\label{fig6}
\end{figure*}

We show the cholesteric version of this structure in Fig.~\ref{fig6}(a,b) at $N=4$. In cholesteric droplets with planar anchoring and sufficiently strong chirality ($N \geq 1$) the Frank--Pryce texture is the lowest energy and most commonly observed structure, whereas the flat layers dominate at smaller $N \leq 1$~\cite{bouligand_organization_1984, sec_geometrical_2012}. 

A similar situation arises in twist-bend droplets. At $M=5$ and $N < 10$, the equilibrium texture resulting from an initial Frank--Pryce structure is very similar to the `spirals' that arise for an initial condition of flat layers, as shown in Fig.~\ref{fig4}. The supports the notion that the twist-bend analogue of flat layers is energetically favoured for smaller $N$, just as is the case in a cholesteric. For larger $N$---we show $N=12$ in Fig.~\ref{fig6}(c,d)---the Frank--Pryce texture is stable and has a structure that is visually very similar to the cholesteric texture. However, the material nucleates additional defects beyond the single $+1$ defect required by topology. The defects in the structure shown in Fig.~\ref{fig6}(c,d) all have either MI 1 (for $-1$ charge, purple) or MI 2 (for $+1$ charge, yellow).

Comparing with Fig.~\ref{fig2}(d,e), we recognise that at large $N$ an initial condition without layers will also equilibrate at an analogue of the Frank--Pryce texture, which we have called the onion, again supporting the idea that this structure is energetically favourable at high $N$. However, just as with the cholesteric case~\cite{sec_geometrical_2012}, flat and cylindrical layers are still metastable states at large $N$. We also expect a diversity of textures with similar structures but differing numbers and MIs of the defects, rather than a single unique version of the onion or Frank--Pryce texture. In Supplemental Movies 10 and 11 we show the development of this texture at $M=5,N=6$ and $M=5, N=12$ respectively. 

For larger values of $M$ the internal structure of the equilibrium textures is very different from the Frank--Pryce structure, and there are no visible spherical layers even at large $N$. The integral curves of the director field form helices that spiral out from a MI 0 defect in the interior of the droplet to the boundary, and they do not have the U-shaped structure that occurs in small $M$ textures such as that shown in Fig.~\ref{fig6}(c,d). We show a selection of such structures in Figs.~\ref{fig6}(e-h), and in Supplemental Movie 12 we show the development of such a texture from an initial Frank--Pryce structure at $M=11, N=12$. These structures are very similar to the textures we have already discussed for large $M$, but with different configurations of the $\beta$-lines. Indeed, the characteristic feature of the textures shown in Figs.~\ref{fig6}(e-h) is the twisted helix of $\beta$-lines emanating from the point defect. Even for large $M$ we do not see the same `pinwheel' PT surface we see of flat layers, Fig.~\ref{fig4}(g,h). As we will see presently in the quenches, this type of structure appears to be the most common metastable structure for any value of $N$.

\subsection{Defect Strings}
\begin{figure*}[t]
\centering
\includegraphics[width=0.98\textwidth]{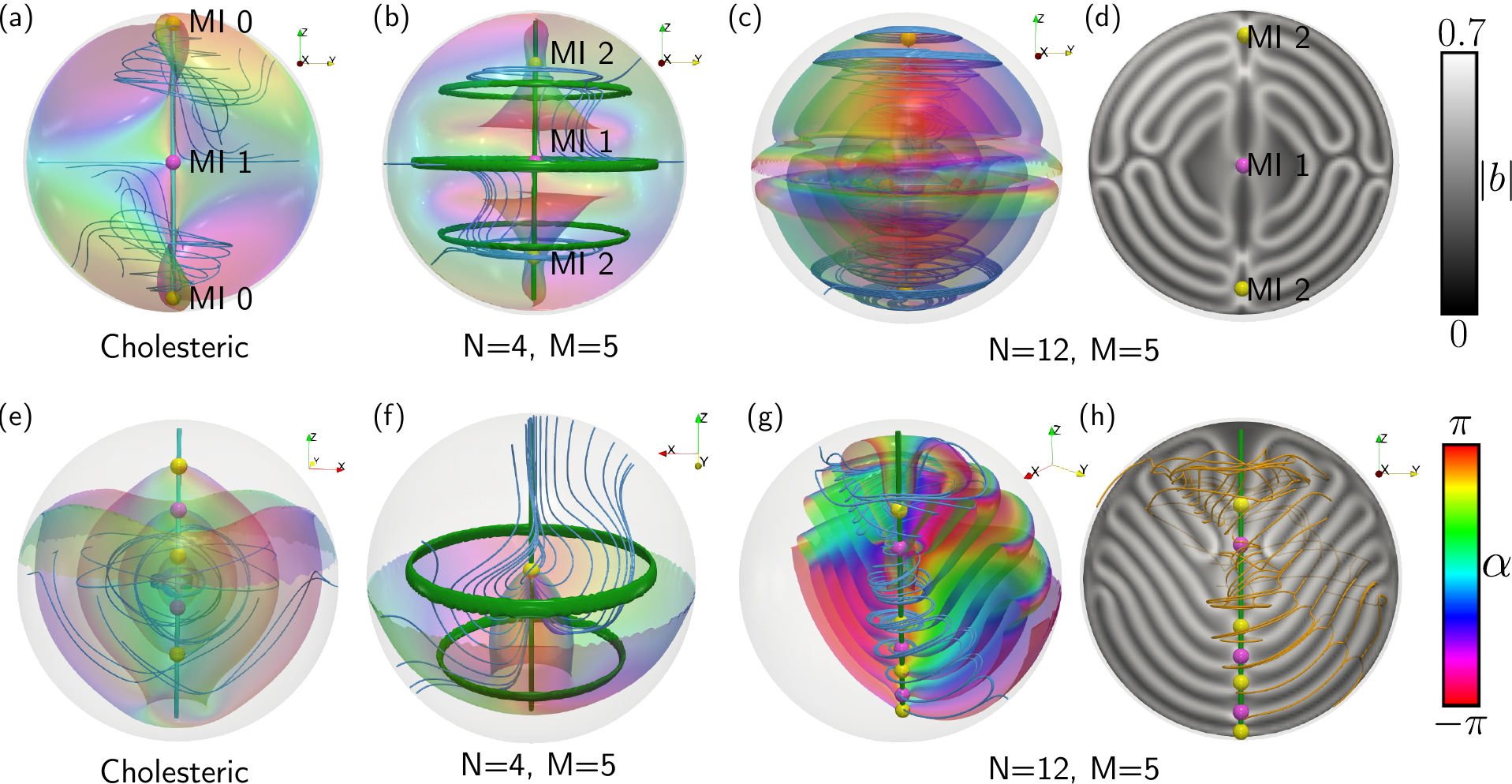}
\caption{Structures containing strings of defects. (a) A stable defect string in a cholesteric droplet at $N=2$. The structure consists of a pair of MI 0 defects bound to the droplet surface, and a central MI 1 defect. A $\lambda$-line connects them. (b) Initialising this structure in a twist-bend nematic droplet at $N=4, M-5$ results in a similar, but different structure. The central charge $-1$ point defect still has MI 1, the others are initially hedgehogs (MI 0) but convert to hyperbolic (MI 2) in the equilibrium structure. In addition to the $\beta$-line connecting the defects, there are three circular $\beta$ lines which occur because of the interpolation between the hyperbolic local structure of the defects and the radial director on the boundary. (c,d) At $N=12, M=5$, the string of three defects remains stable but the droplet fills with nested layers of the `Mexican hat' type. The $\beta$-lines (not shown), form a dense, complex network. In panel (d), we show the magnitude of the bend vector field across a slice, which helps visualise the layer structure. (e) Nested spherical layers in a cholesteric droplet at $N=3$. Each spherical layer is a toron---the droplet contains two torons, plus an additional boundary defect, for a total of $N$. (f) Initialising this structure in a twist-bend nematic droplet (here, $N=4, M=5$) produces an `octopus'-like structure with a central MI 2 point defect. (g,h) At larger $N=12$ the string of $N+1$ point defects is stable but the layers are not spherical---rather, they resemble the `flower' structure. For clarity we have shown only half of the PT surface, the half with $x < 0$---the removed half is the mirror image of this. In each panel the defect positions are indicated with a yellow sphere for charge $+1$ and a magenta sphere for charge $-1$. The MI is as indicated. The coloured surface is the PT surface with $n_z=0$, coloured according to the angle $\alpha = \tan^{-1}(n_y/n_x)$. Integral curves of the director are shown in blue, the bend in orange. In panels (b,f) the $\beta$-lines are indicated by green tubes---we omit the $\beta$-lines in other panels for clarity. In panels (d,h) the magnitude of the bend is shown on a slice.}
\label{fig7}
\end{figure*}

Structures comprised of `strings' of point defects, connected by $\lambda$-lines, occur at wide range of chiralities $N \geq 1$ in cholesteric droplets, with longer strings (i.e., containing more defects) becoming more common at higher chirality~\cite{posnjak_hidden_2017, posnjak_topological_2018, pollard_point_2019}. The simplest such string occurring in a cholesteric droplet consists of a pair of charge $+1$ hedgehog point defects (MI 0) pinned to the boundary, with a central charge $-1$ hyperbolic point defect (MI 1) in the centre. This is shown in Fig.~\ref{fig7}(a). More complex strings involve nested spherical layers, the inclusion of torons~\cite{wu_hopfions_2022} (essentially an isolated spherical layer) into other structures, and the decomposition of high-charge defects~\cite{posnjak_topological_2018}.

We initialise the twist-bend nematic with this cholesteric equilibrium director, simply by inserting a MI 1 defect at the centre of the droplet and then imposing the radial boundary condition. At small $N < 4$ the resulting string is not stable, and the boundary point defects move in and annihilate with the defect in the centre, producing an equilibrium structure with a twisted hedgehog in the centre of the droplet, as in Fig.~\ref{fig1}(b). At larger values of $N$ the boundary defects change their structure from MI 0 to MI 2, as eventually occurs for the twisted hedgehog defect. While they move away from the boundary as part of this process, they do not annihilate with the central defect and we still have a stable string of three point defects, Fig.~\ref{fig7}(b), albeit one which is different in structure to the cholesteric texture shown in Fig.~\ref{fig7}(a). There is a $\beta$-line along the axis containing the defects, and three additional $\beta$-line loops encircling it. We show the development of this structure in Supplemental Movie 13, for parameter values $M=5, N=4$. 

Making $N$ larger still, the droplet fills with `Mexican hat' layers, as in the case of the single central defect, but the other two defects remain. The integral curves of the director field are not U-shaped, rather they are tight spirals, almost circles. For larger $M$ the string decays to a radial hedgehog for a wider range of $N$ values, but at large $N \approx 12$ the defect string remains, and the droplet is filled with a complex, twisted network of $\beta$-lines. 

Spherical layers are also a common structure in cholesteric droplets with both planar and radial anchoring. If a director field is tangent to a sphere, then it must have singularities with a total charge of $+2$. In the absence of disclinations, this means a single $+2$ singularity, or a pair of $+1$ singularities. Initialising a director tangent to concentric spheres with a $+2$ singular line produces the Frank--Pryce structure. The alternative is to have a $\chi^{+1}$-line along an axis of the droplet. To maintain chirality, this will `escape' alternatively up and down to remove the singular line and produce a string of point defects~\cite{sec_geometrical_2012, pollard_escape_2024}. These point defects sit on spherical layers, called torons in cholesterics~\cite{posnjak_points_2016, wu_hopfions_2022}, so the structure is a collection of nested torons, which occur in cholesteric droplets at a wide range of chiralities~\cite{posnjak_topological_2018}. With planar anchoring, each point defect will be hyperbolic, alternatively MI 1 (charge $-1$) and MI 2 (charge $+1$); in a droplet with radial anchoring there will be an additional boundary hedgehog. An example of such a texture, in a cholesteric at $N=3$, is shown in Fig.~\ref{fig7}(e).

Spherical layers with a $\chi$-line along the $z$-axis are initialised using the director field,
\begin{equation}
    {\bf n} = \cos(2N\pi r/R){\bf e}_\theta + \sin(2N\pi r/R){\bf e}_\phi.
\end{equation}
At small $N$ the strings are not stable in a twist-bend nematics, and decay to  a state with a central defect that is a twisted hedgehog, shown in Fig.~\ref{fig7}(f) for $N=4, M=5$. The PT surface again resembles an octopus. At larger $N$ we do indeed reproduce a string of defects, even though it is possible to `escape' the singular line without producing point defects when there is no requirement for chirality. We show the equilibrium texture for $N=12, M=5$ in Fig.~\ref{fig7}(g,h). In contrast to the cholesteric, the twist-bend droplet is filled not with spherical layers, but flower-shaped layers as in Fig.~\ref{fig5}(e,f), with the typical U-shaped integral curves for the director field.

\begin{figure*}[t]
\centering
\includegraphics[width=0.98\textwidth]{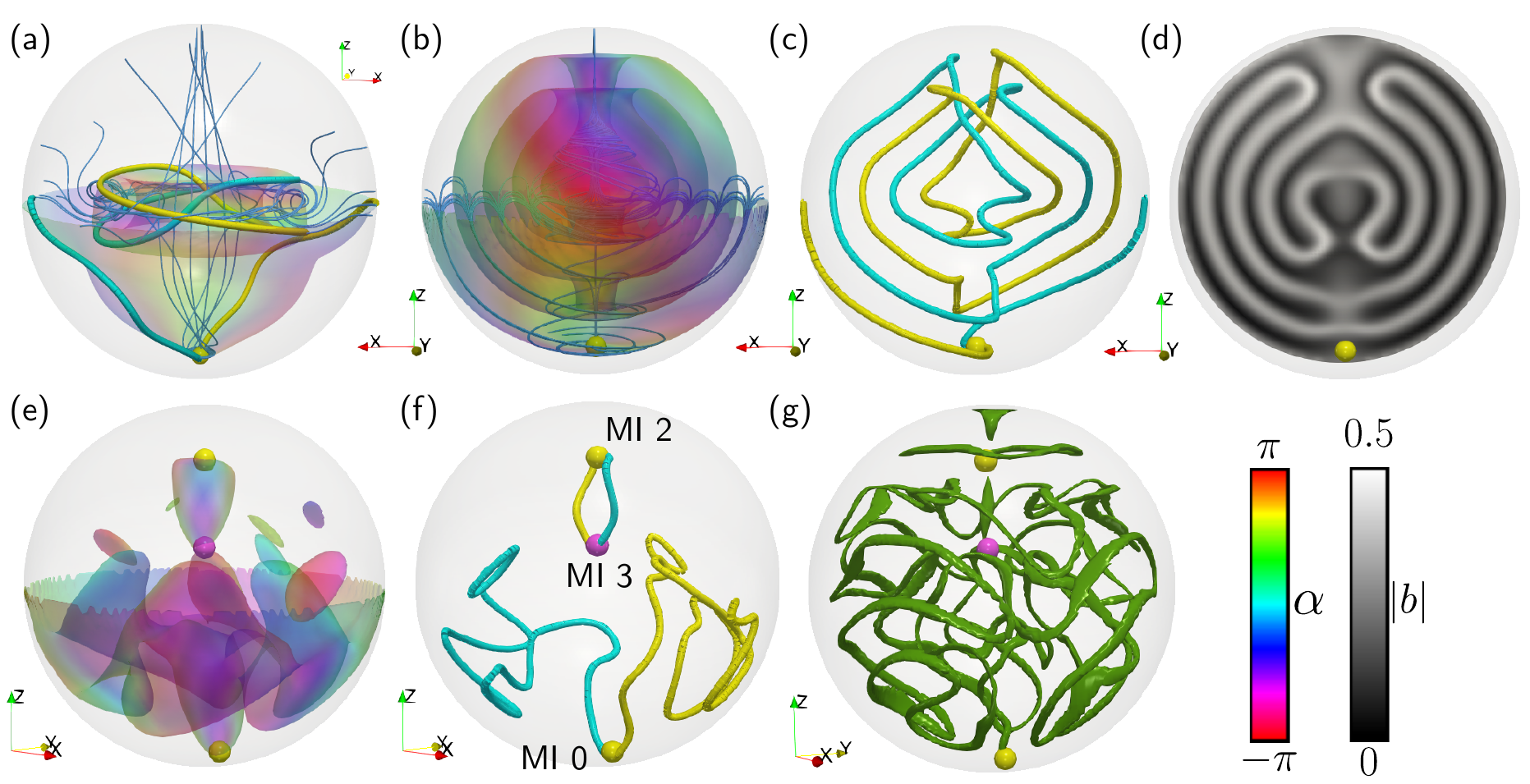}
\caption{The Lyre structure. (a) The classical Lyre structure in a cholesteric droplet, at $N=4$. The preimages of $-{\bf e}_x$ (blue tube) and $+{\bf e}_x$ (yellow tube) are linked, illustrating the nontrivial Hopf invariant. (b-d) Analogous structure in twist-bend nematic droplet, at $N=10, M=5$. (b) The PT surface with $n_z=0$, coloured according to the angle $\alpha = \tan^{-1}(n_y/n_x)$. Integral curves of the director are shown in blue. (c) The preimages of $-{\bf e}_x$ (blue tube) and $+{\bf e}_x$ (yellow tube) are linked as in the cholesteric texture. (d) The magnitude of the bend is shown on a slice. (e-g) At $N=10, M=6$ the Hopfion eventually decays, being replaced by a toron. (e) This toron is visible as a spherical component to the PT surface, on which a pair of point defects of opposite charge sit. Unlike in the cholesteric toron, the defects have Morse indices (MI) 3 (purple, $-1$ charge) and 2 (yellow, $+1$ charge). The boundary defect in this texture, as well as those shown in (a) and (b-d), is MI 0. (f) There is no linking of preimages, illustrating the Hopf invariant is zero. (g) A dense network of $\beta$-lines (green) fills the droplet. }
\label{fig8}
\end{figure*}

\subsection{High Charge Defects}
Cholesteric droplets with high chirality exhibit point defects and disclination lines with defect charge $|q| > 1$~\cite{posnjak_hidden_2017, posnjak_topological_2018, pollard_point_2019}. Notably, there are point defects with charge $-2$ and a triangular symmetry, as well as point defects with charge $-3$ with a tetrahedral symmetry. These structures are stabilised by the boundary hedgehog defects, which are pinned to the boundary and rearrange themselves to achieve maximal separation---this involves forming an equilateral triangle in the case of three boundary defects, and a tetrahedron in the case of four boundary defects. If the symmetry is broken then these high-charge defects may also split apart into multiple defects of lower charge, creating long defect strings. The possible ways in which this may occur are predicted by singularity theory~\cite{pollard_point_2019}.

We simulate twist-bend nematic droplets containing charge $-2$ and $-3$ defects using initial directors obtain from singularity theory---see Appendix~\ref{appA}---which for cholesterics qualitatively reproduce the equilibrium structures observed in experiments. For all parameter values surveyed, the high-charge defects are not stable and immediately decompose into multiple generic defects of winding $-1$. At small $N$, the resulting equilibrium textures are twisted hedgehogs, as in Fig.~\ref{fig1}(b). At larger values of $N$ we stabilise longer strings of point defects. 

A $-2$ defect splits into a pair of $-1$ defects with MI 1. The three boundary defects are hedgehogs, MI 0, and the resulting structure forms a V-shaped string of five defects which is stable in a cholesteric~\cite{pollard_point_2019}. In a twist-bend nematic these strings are also stable for intermediate $N$. A complex network of deformed $\beta$-lines connects the defects in the string, as well as filling the rest of the droplet. This is quite different from the cholesteric case, where the $\lambda$-lines are the core of straight double-twist cylinders connecting the defects in the string~\cite{pollard_point_2019}. Strings with a similar structure are observed in our quenches. At larger cone angles, the equilibrium structure at small $N$ is the central hyperbolic defect shown in Fig.~\ref{fig1}(c), while at high $N$ the defect string remains, but the droplet is now filled with pseudolayers.

The lack of stability of the central high-charge defects here is not surprising. It results from a combination of that fact that (i) hedgehog defects are not bound to the droplet surface as they would be in a cholesteric, and (ii) the cost of creating regions of reversed handedness is significantly lower in a twist-bend nematics; together, these considerations imply a freedom of movement that allows for relaxation pathways not accessible to the chiral material. 

Defects with a high {\it positive} charge have not been observed in cholesteric droplets. This is because it is the surface-bound positive charges that stabilise the negatively-charged central defect, whereas positive central charges would require additional negatively charged defects which are not bound to the surface. For example, a central $+2$ defect would require, in the minimal case, an additional MI 1 defect with charge $-1$ elsewhere in the droplet to meet the constraints of the Poincar\'e--Hopf theorem. Defects with $+2$ charge defects have a large bend distortion and U-shaped integral curves similar to those observed in the other textures we have examined, so it is reasonable to wonder whether they might be stable in a twist-bend nematic. However, such defects were not stable in our simulations at any parameter value we examined. At smaller values of $N \leq 4$ the equilibrium structure is either a twisted hedgehog or a central hyperbolic defect depending on the value of $M$, while for large $N$ a $+2$ defect splits into a pair of defects, one with MI 0 and one with MI 2, that form a stable string with the boundary MI 1 defect.

\section{Hopfions and Torons in Twist-Bend Droplets}
Hopfions, and analogous structures such as heliknotons, are three-dimensional solitons consisting of knotted or linked soliton tubes---see Appendices~\ref{appB} and~\ref{appC} for more details. In an achiral nematic Hopfions will shrink and eventually decay to a uniform state, but they can be stabilised in bulk chiral materials~\cite{chen_generating_2013, ackerman_diversity_2017, tai_three-dimensional_2019, wu_hopfions_2022}. The classical cholesteric Hopfion consists of a pair of linked $\lambda$-lines that are the cores of double-twist cylinders. Hopfions also occur in cholesteric droplets with both planar and radial anchoring in the form of the Lyre and Yeti structures~\cite{sec_geometrical_2012, posnjak_topological_2018}. These structures are topologically the same, but have slightly different director fields, with the Lyre being preferred when the anchoring is radial. Phenomenologically, these structures involve a mixture of cylindrical and toroidal layers. They contain a $+1$-winding $\lambda$-line connecting opposite poles on the sphere---along the $z$-axis, say, an enclosed by cylindrical layers---as well as a second $\lambda$-line in the form of a closed loop encircling the first, enclosed by toroidal layers that fill the part of the droplet not filled by the cylindrical layers. This linking (relative to the boundary) of $\lambda$-lines ensures a nonzero Hopf invariant as detailed in Ref.~\cite{binysh_geometry_2020}. The more familiar way of computing the Hopf invariant is from a linking of preimages of the director, as explained in Appendix~\ref{appB}.

The Derrick--Hobart theorem~\cite{derrick_comments_1964} guarantees that, in the absence of a stabilising lengthscale, shrinking a Hopf soliton will always reduce the energy, and hence these features are never stable in an ordinary achiral nematic. They can be stabilised in cholesterics precisely because of the lengthscale arising from the pitch length. However, in addition to this shrinking there is a second mechanism by which Hopf solitons may be destroyed. As a Hopf soliton essentially involves the linking of two closed meron tubes, we may remove the soliton by `unwinding' the meron tube via the creation of a pair of point defects. These point defects may later annihilate again, leaving a (topologically) uniform texture behind, or a material lengthscale may stabilise them at some separation, leaving behind a feature called a toron---a pair of point defects which sit at opposite poles of a spherical component of a PT surface and are connected by a (topologically required) $\lambda$-line or $\beta$-line. For a toron arising from the decomposition of a Hopfion, one of these point defects must be a hedgehog (MI 0 or MI 3) and, as we have discussed, these defects are incompatible with a consistent sense of handedness; the strong energetic requirement for maintaining a consistent sense of handedness in a chiral material therefore also helps contribute to the stability of Hopfions by making this mechanism of decay less favourable. 

Like cholesterics, twist-bend nematics also have an inherent lengthscale that allows us to violate the Derrick--Hobart theorem. However, while twist-bend nematics favour nonzero twist there is no energetic requirement that the sense of handedness be consistent, and as we have seen in the simulations presented in this work they are consequently more accommodating of hedgehog defects. Thus, it is not \textit{a priori} obvious whether or not Hopfions will be stable in these materials. 

We initialise the Lyre structure in a droplet using the director field,
\begin{equation}
    {\bf n} =\cos(\pi\tau/\tau_0){\bf e}_\phi + \sin(\pi\tau/\tau_0){\bf e}_\psi,
\end{equation}
inside a tube with $\tau \leq \tau_0$, and taking ${\bf n} = {\bf e}_z$ throughout the rest of the droplet. Here, $\rho = \sqrt{x^2+y^2}$, $\tau = \sqrt{(\rho-\rho_0)^2 + z^2}$, and $\psi = \tan^{-1}(z/(\rho-\rho_0))$. This initialises double-twist cylinders along the curves $\rho=0$ and $\rho=\rho_0$. We take the parameters $\rho_0 = 0.6R$, $\tau_0 = 0.2R$, but this choice is not important for the equilibrium structure. 

We show the cholesteric Lyre structure at $N=4$ in Fig.~\ref{fig8}(a). The set of curves where the director is aligned with $\pm {\bf e}_x$ (yellow and blue tubes) are linked, and this illustrates the nontrivial Hopf invariant of $+1$. This can also be seen from the $\lambda$-lines, which in the cholesteric texture consist of the $z$-axis, along with a second loop at the core of the toroidal region (essentially the curve $\tau =0$) enclosing this. 

To simulate Hopfions in twist-bend nematics we begin from the same Lyre initial condition that is close to being the equilibrium state in a cholesteric. For small $N$ and for all values of $M$ the Hopfions are unstable, and rapidly decay to a bowl-like state. For larger cone angles $M = 5, 5.5$ there is a narrow window of values of $N$ for which Hopfions are stable, approximately $N=6-10$ for $M=5$ and $N = 10-12$ for $M=5.5$. An equilibrium texture containing a Hopfion at $M=5, N=10$ is shown in Fig.~\ref{fig8}(b-d), with the evolution from the initial Lyre structure shown in Supplemental Movie 14. In Supplemental Movie 15 we show the evolution of a Hopfion at $M=5.5, N=12$. Structurally, these textures resemble a mix between the onion and the flower textures, quite different from the Lyre texture that we use as the initial condition. For larger values of $N$, Hopfions again become unstable, with the large $N$ textures being reminiscent of the flower texture with additional point defects. 

The means by which Hopfions decay in our simulations is the second of the two processes identified above: the two preimages move closer together until they finally `intersect', which corresponds to the creation of a pair of point defects. This occurs, in part, because the integral curves of the director are coiling up into jagged helices, which can be seen in the simulation shown in Supplemental Movie 16 for $M=6, N=8$. Such a process destroys the Hopfion and replaces it with a toron. At small $N$ and/or large $M$ the toron coarsens away, but at intermediate values of $N,M$ the toron remains a stable feature of the equilibrium textures. For example, for $M=6$, $N=8-12$ the toron remains---we show $M=6, N=10$ in Fig.~\ref{fig8}(e-g)---while for $M=7$ we only observed the toron being stable at $N=12$, and we did not find stable torons for any $N \leq 12$ at $M > 7$. For smaller $M = 5 -7$ Hopfions take a significantly longer time to reach the final equilibrium state than the other initial conditions discussed in this work. 

The toron in our twist-bend nematic differs from the usual cholesteric toron in that it consists of a MI 3 point defect and a MI 2 point defect, rather than a MI 2 and MI 1 pair. Even in situations where the torons decay---by pairwise annihilation of the point defects---we find they persist for a long simulation time, much like the Hopfions themselves. It is therefore possible that both torons and Hopfions may be observed as a transitory metastable feature in twist-bend droplets, even at parameter values where they are not part of the equilibrium textures. 

We also find that, unlike the other textures examined in this work, the behaviour of Hopfions is quite sensitive to numerical parameters such as the number of points used in the discretisation of the droplet, with too few simulation points resulting in the sharp `kinks' in the director field not being well approximated, leading to numerical instabilities. The energy landscape for twist-bend nematics is evidently quite complex, and we suspect that Hopfions correspond either to a very narrow minimum in the energy, or else lie close to a slow trajectory past a saddle in the energy, making them metastable for long times. Anisotropy among the elastic constants also plays a role in the stability of Hopfions in cholesterics. Our simulations only cover the case of equal elastic constants---it may well be that a realistic choice of elastic constants enlarges the parameter regime in which they are stable. It would therefore be interesting to perform a more detailed numerical and analytic study of Hopfions in bulk bent-core materials, perhaps carried out alongside experimental studies applying the techniques outlined in Refs.~\cite{chen_generating_2013, tai_three-dimensional_2019, wu_hopfions_2022} to generate the solitons.

\section{Stable Structures Arising From Quenches}
Now that we have surveyed some of the stable structures we can expect to observe in twist-bend nematics, we estimate their prevalence by simulating quenches from the isotropic to the twist-bend nematic phase and examining the statistics of the stable structures that result from this quench. Accordingly, for each set of parameter values we initialise 100 simulations with random initial conditions, and then minimise the free energy until we achieve equilibrium. 

\begin{figure*}[t]
\centering
\includegraphics[width=0.8\textwidth]{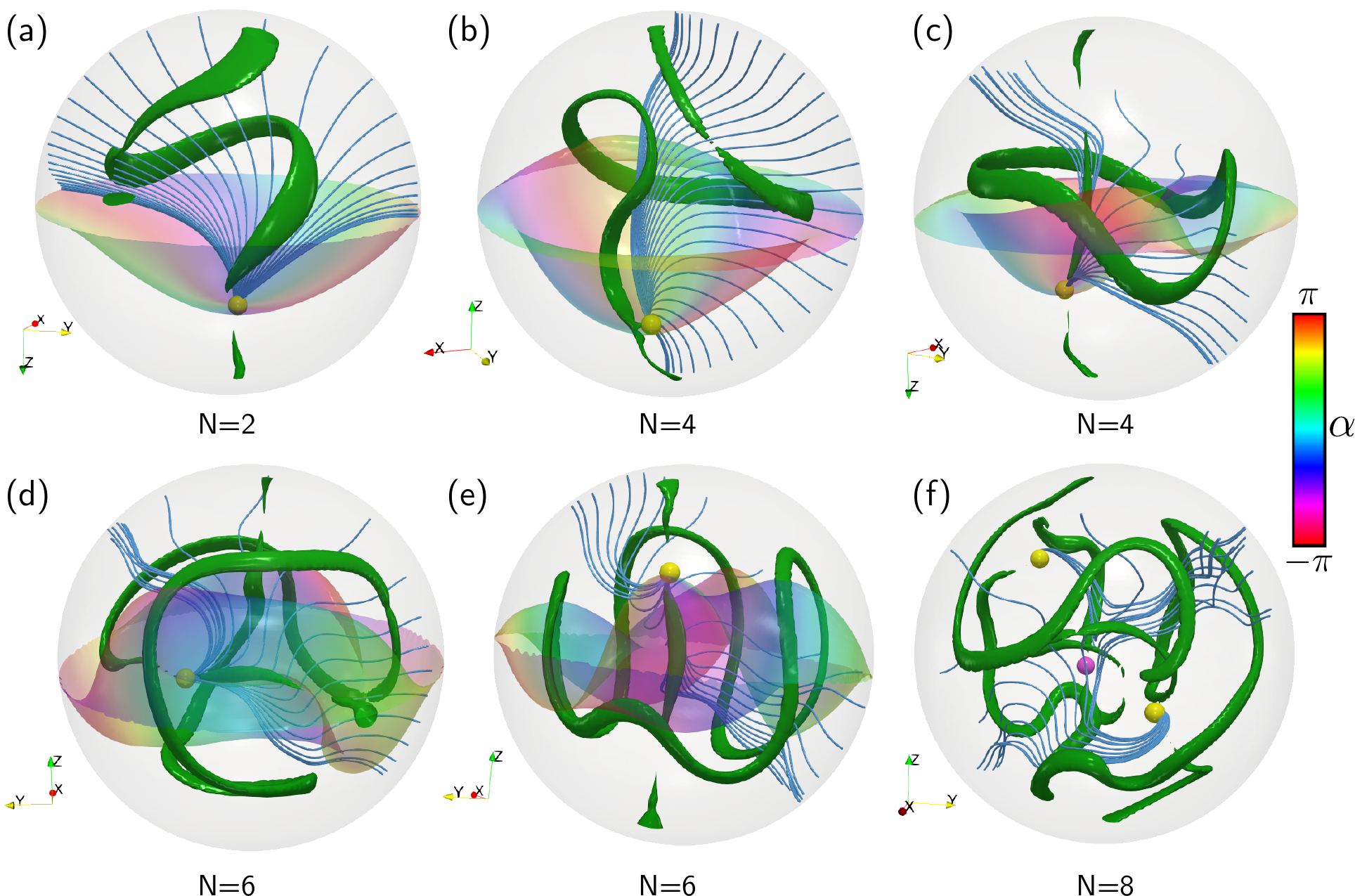}
\caption{Equilibrium structures resulting from a quench in a twist-bend nematic at $M=7.5$ and at various values of $N$. The surfaces are PT surfaces with $n_z=0$, coloured according to the angle $\alpha = \tan^{-1}(n_y/n_x)$. The bend zeros are shown as green tubes. Point defects are indicated by yellow ($+1$ charge, MI 0 in all examples) and magenta ($-1$ charge, MI 1) spheres. Integral curves of the director are shown in blue.}
\label{fig9}
\end{figure*}

At smaller values of $N$ the equilibrium textures arising from these quenches all contain either a single central or boundary defect, more commonly the later, and are distorted versions of either the `bowl' textures, or the textures with a central defect. At intermediate $N$, the variants on the cholesteric `flat layer' and Frank--Pryce structures dominate, that is, the octopus and screw textures---this is similar to the cholesteric, where flat and spherical layers are the most common structures~\cite{sec_geometrical_2012}. At larger $N$ we see variants on the flower, onion, and pinwheel structures, often with multiple defects. There is some disparity in the number and geometry of $\beta$-lines arising in these textures, which we describe presently for a specific value of $M$. We did not observe a single Hopfion texture at any parameter value examined in our simulations, suggesting that such structures are rare and will occur with less than $1\%$ frequency in a quench. 

We limit ourselves to a detailed analysis of the case of $M=7.5$, an intermediate value in our parameter regime which corresponds to the cone angle for the twist-bend phase reported by Chen et. al. in the molecule CB(CH$_2$)$_7$CB~\cite{chen_chiral_2013}. The pitch length of this material is $8.3\text{nm}$, and we survey droplets with radius ranging from $16.6\text{nm}$ ($N=2$) to $99.6\text{nm}$ ($N=12$). 

At $N \leq 4$ the only structures observed contain a single hedgehog defect, MI 0, which is either close to the droplet centre or to the droplet boundary. These are reminiscent of the `bowl' structure. Most contain a single $\beta$-line, twisted up into a helix, on which the point defect sits. At $N=4$ 19\% of the structures have a second $\beta$-line in the form of a closed loop, a result of the single, twisted $\beta$-line intersecting itself and `pinching off'. We show representative structures in Fig.~\ref{fig9}(a,b,c).

At $N=6$ we still see only single defects with MI 0. Most structures have a single $\beta$-line with a helical structure, as in Fig.~\ref{fig9}(d). In the majority of the 31\% of structures with a second, closed loop $\beta$-line, it has the same undulating structure as in the `pinwheel' texture, Fig.~\ref{fig9}(e). This trend continues at $N=8$, where 43\% of a structures had a second, closed loop $\beta$-line, and 4\% had two closed loop $\beta$-lines. Additionally, at this value of $N$ strings of three defects (two MI 0, one MI 1) begin to appear rarely, Fig.~\ref{fig9}(f), with 2\% of the simulations showing this structure---the remaining 98\% had a single MI 0 point defect. At $N=10$ strings of three defects become slightly more common, comprising 9\% of the structures observed. At $N=12$ all structures are clearly of the pinwheel type, with at least one closed loop, undulating $\beta$-line. 23\% contain a string of three defects (two MI 0, one MI 1), with the remainder having only a single MI 0 defect. None of the extra defect pairs form a toron. This trend continues for larger $N$, with longer defect strings becoming more common with increasing $N$, as in cholesterics~\cite{posnjak_topological_2018}.

\section{Discussion}
In this work we have surveyed the complex textures that can arise in spherical droplets of a twist-bend nematic material with radial anchoring. We have shown that many of the equilibrium structures are analogous to those found experimentally in cholesteric droplets with radial anchoring, but with quite different local structures. The most common structures that arise contain either a central point defect with a twisted structure, or are analogous to the `flat layer' and Frank--Pryce cholesteric textures, but defect strings and textures with cylindrical layers are also metastable. The ability to vary not just the pitch length, but also the cone angle, leads to an even wider variety of topologically and geometrically distinct structures than are seen in cholesterics, and it is clear that our study has only scratched the surface of what is possible with these materials and the many different kinds of confinement they can be subjected to. Typically, we find that stability of the more interesting structures requires both a large cone angle, as well as significantly larger values of the dimensionless parameter $N$ than in a cholesteric, although the shorter pitch lengths of twist-bend materials mean that achieving these values of $N$ does not actually require very large droplets.

In particular, we have found Hopfions to be numerically stable for a narrow parameter regime. Although it is possible that in a real material Hopfions could stabilised at other values of the parameters due to effects we have not considered here. For example, anisotropy among the elastic constants could have a stabilising effect. Furthermore, we have only studied a particular initial structure that is suited to the cholesteric case, and it may be that the geometric structure of Hopfions needs to be substantially different in twist-bend nematics in order for them to be stabilised at a wider range of parameter values. The substantial change that the director field undergoes as we transition from the initial to the equilibrium state tends to support this idea.  

Trying to pack helical filaments into a sphere inevitably results in frustration, and we have observed several ways in which the material tries to deal with this. At large cone angles two characteristic motifs emerge in the helical integral curves of the director. Either they become very tight spirals that are almost flat circles, or they are U-shaped, being almost straight but with a sudden tight bend. At small cone angles the integral curves almost become flattened and circle-like, but with a pronounced helical wobble. In both cases the bend distortion is very frustrated and there are complex networks of $\beta$-lines, zeros in the bend field. These $\beta$-lines are not generally associated to meron/Skyrmion tubes, as the $\lambda$-lines in a cholesteric typically are---rather, they are inflection points in the integral curves resulting from the geometric frustration of the packing. Although we have not analysed this frustration in detail here, we have noted the similarity between aspects of the observed structures and models of hyperbolic geometry, similarly to how cholesterics exhibit aspects of spherical geometry. Using methods of differential geometry to study the ways in which helical fibres can be packed into spheres with radial boundary conditions would be interesting, and perhaps help to shed further light on these structures. Several recent works on geometric frustration in modulated phases of nematics provide a helpful framework for such investigations~\cite{niv_geometric_2018, sadoc_liquid_2020, kamien_geodesic_2020, chaturvedi_gnomonious_2020, binysh_geometry_2020, pollard_intrinsic_2021, da_silva_moving_2021, selinger_director_2022}. These methods may also help analyse the question of Hopfion stability, by providing director field structures that are more adapted to the twist-bend material, rather than the cholesteric. 

There is scope for extending this study to consider different kinds of anchoring conditions or different kinds of confinement. It is natural to consider planar anchoring on a spherical droplet. Besides a few `special' structures with high-charge defects that are unique to the case of radial anchoring, in cholesteric droplets the structures are much the same between the two different anchoring conditions, and any structure in a droplet with planar anchoring can be extended to a structure in a droplet with radial anchoring by the addition of a `boundary layer'~\cite{pollard_topology_2020}. We suggest that a similar situation arises for twist-bend nematics, where radial anchoring is again the more frustrating of the two boundary conditions. Future studies could also explore the effects of elastic anisotropy on the observed structures, alternative couplings between the director and polarisation field---for example, a coupling that promotes splay distortion instead of bend, which would naturally be more frustrated by planar rather than radial anchoring---or other kinds of molecular geometry entirely, such as tetrahedral molecules~\cite{rosseto_modulated_2022}. Cone angles and pitch lengths of bent core materials can be temperature dependent~\cite{jakli_physics_2018}, so it may also be useful to study how an equilibrium structure responds to a sudden change in these material parameters---the equilibrium textures themselves are quite complex, and the transition pathways between them are also likely to be interesting. 

\appendix

\section{Point Defects and Singularity Theory}
\label{appA}
A generic point defect is described by a director field ${\bf n} = \nabla \phi/|\nabla \phi|$, where 
\begin{equation} \label{eq:morse}
    \phi = \frac{1}{2}\left(\pm x^2 \pm y^2 \pm z^2 \right).
\end{equation}
Each $\pm$ sign can be chosen independently in this formula; the total number of $-$ signs is the Morse index (MI) of the singularity. MI 0 and 3 are `hedgehog' defects, MI 1 and 2 are `hyperbolic'; the defect charge is $+1$ if the MI is even and $-1$ if the MI is odd. Only MI 1 and 2 defects can be made `chiral', with twist ${\bf n} \cdot \nabla \times {\bf n} \neq 0$ in a neighbourhood of the defect~\cite{pollard_point_2019}, and thus these are energetically preferred in a cholesteric. The passage from MI 0 to MI 2 can be effected by a homotopy of the director field, and it necessarily involves a `Hopf bifurcation'~\cite{pollard_morse_2024} which produces a twisted structure with an `inversion ring'~\cite{ciuchi_inversion_2024}. 

Higher-charge defects can be described by their diffeomorphism class, and given a symmetry-based label using the machinery of singularity theory, which also provides model director fields~\cite{arnold_singularity_1998, pollard_point_2019}. In a cholesteric droplet the singularity classes that show up are those that are favoured by a maximal separation of the boundary defects, which sets a global symmetry group of the droplet~\cite{pollard_point_2019}. The charge $-2$ defect in a cholesteric droplet is described locally by a director of the form ${\bf n} = \nabla \phi/|\nabla \phi|$, where $\phi$ is the $D_4^-$ singularity,
\begin{equation} \label{eq:d4}
    \phi = x^2y - \frac{1}{3}y^3 + \frac{1}{2}z^2
\end{equation}
Flipping the sign of any of the components of the director will produce a charge $+2$ singularity. The charge $-3$ defect has a tetrahedral symmetry, described locally by the $T_{444}$ singularity,
\begin{equation} \label{eq:t444}
    \phi = \frac{1}{4}(x^4+y^4+z^4) + axyz,
\end{equation}
where $a$ is a parameter with dimensions of length, which we take to be $R/\sqrt{3}$.

\section{Topological Solitons}
\label{appB}
Solitons are nonsingular---but still topologically protected---structures that can appear within nematic textures, as well as other oriented media such as ferromagnets. Mathematically, solitons are associated with two topological invariants of the director fields. 

The first, the Euler class, can be evaluated on a surface to give the Skyrmion charge across that surface~\cite{pollard_morse_2024, machon_global_2016}. Physically, the Euler class corresponds to line-like meron and Skyrmion tubes (double-twist cylinders in cholesterics) as well as edge- and screw-dislocations. While we do not make use of this invariant in this work, for completeness we note that the value of the Euler class across an oriented surface $\Sigma$ is $2Q$, where $Q$ is the Skyrmion charge of $\Sigma$, and it can be determined by the simple formula~\cite{copar_topological_2012, pollard_morse_2024}
\begin{equation} \label{eq:skyrmion}
    2Q = \chi(\Sigma^+) - \chi(\Sigma^-),
\end{equation}
where $\chi$ denotes the Euler characteristic and $\Sigma^+, \Sigma^-$ are the parts of the surface where the director points out and in, respectively. When $\Sigma$ is a sphere enclosing a point defect then $Q$ is the defect charge---generically point defects sit at the confluence of meron lines. Thus, we can compute the Skyrmion charge across $\Sigma$ by orienting the director field and examining the dot product between the director and the surface $\Sigma$ visually. Since this same construction can be used to visualise the pseudolayer structure within the material, this also helps to provide a topological meaning to the pseudolayers. 

The second topological invariant, the Hopf invariant, is computed from the linking number of a pair of preimages of the director field---e.g., the set of points where ${\bf n} = \pm {\bf e}_x$---which are generically lines~\cite{chen_generating_2013}. The Hopf invariant is associated with fundamentally three-dimensional solitons compressed of knotted or linked tubes of merons/Skyrmions, often called Hopfions or heliknotons~\cite{chen_generating_2013, ackerman_diversity_2017, tai_three-dimensional_2019, wu_hopfions_2022}. We identify these in our simulations by looking at the lines where ${\bf n} = \pm {\bf e}_x$ and computing the linking number. For the textures shown in this manuscript these lines are simple enough that this linking number can be computed by inspection. 

\section{$\lambda$- and $\beta$-lines}
\label{appC}
As well as topological structure, solitons convey geometric structure within the material. This geometric interpretation is obtained by relating the solitons to the gradients of the director field through the use of auxiliary directions in the material~\cite{machon_umbilic_2016, binysh_geometry_2020}, which further helps us to understand their physical meaning through their relationship with light transmission, as well as their energetic cost. 

A cholesteric has a `pitch axis' that determines its optical properties~\cite{efrati_orientation-dependent_2014, beller_geometry_2014}. The pitch axis can be identified from the perpendicular gradients of the director, which can be decomposed into irreducible representations as in Refs.~\cite{machon_umbilic_2016, selinger_interpretation_2018, pollard_intrinsic_2021, da_silva_moving_2021}. We denote by $\Delta$ the traceless and symmetric part of the gradients. The pitch axis is defined to be the eigendirection of $\Delta$ corresponding to the positive eigenvalue. This direction is necessarily orthogonal to the director, and hence its zeros will be lines and not points. Further, this property endows it with a topological meaning---the Euler class and Hopf invariant can be be computed by a count of the winding and linking numbers of its zero lines, respectively~\cite{machon_umbilic_2016, binysh_geometry_2020}. The eigendirections of $\Delta$ may not be orientable, so the winding around these `$\lambda$-lines' can be any half integer. While we might imagine that $\pm 1/2$ winding would be generic, in fact $+1$-winding $\lambda$-lines are the most common because they sit at the core of `double-twist' cylinders, an energetically favoured structure in a cholesteric and the most common geometric manifestation of the meron/Skyrmion. Because $\lambda$-lines are defects in the optical axis they strongly scatter light, and can therefore be observed in experiments. 

The analogue of $\lambda$-lines in twist-bend materials are the $\beta$-lines, zeros of the bend vector field, which were introduced in Ref.~\cite{binysh_geometry_2020}. The presence of these lines is the signature of a variety of cholesteric- and smectic-like structures including merons and Skyrmions, screw dislocations, twist-grain boundary textures, and focal conics. The key difference between $\lambda$- and $\beta$-lines is that former are zeros of a line field and may have half-integer winding, while bend is a vector field and its zeros can only have integer winding. Nonetheless, because the bend is also a vector field orthogonal to the director the topological properties of $\lambda$- and $\beta$-lines are the same, as detailed in Ref.~\cite{binysh_geometry_2020}. In particular they are topologically constrained to appear in, for example, meron/Skyrmion tubes, but we emphasise that not every $\beta$-line necessarily signals the presence of such a soliton---indeed, in many of our examples the $\beta$-lines are not the core of soliton. 

Finally, we remark on one key difference between $\lambda$- and $\beta$-lines: while $+1$-winding $\lambda$-lines sit at the core of energetically-favoured double-twist regions in a cholesteric, $\beta$-lines pick out the vanishing of the energetically-favoured bend mode in a twist-bend nematic, and are consequently always centres of frustration in the texture which are, at least naively, energetically dis-favourable. 

\section{Numerical Simulations and Visualisation of Textures}
\label{appD}

We simulate structures in twist-bend nematics by numerical minimisation of the free energy in Eq.~\eqref{eq:tb_energy} and in cholesterics by minimisation of Eq.~\eqref{eq:ch_energy}. We minimise the energy using a finite difference scheme on a cubic grid comprised of $100^3$ grid points, starting from either a random initial condition for a quench, or else from a specific initial condition as detailed in the main text. We impose the boundary condition of radial anchoring by fixing the director at all grid points outside a given radius. Simulations are terminated when the change in energy between timesteps falls below a threshold value. We also explored increasing the number of grid points to $150^3$ and $200^3$. This did not make a substantial difference to the textures observed. 

The elastic constant $K$ is the same for all simulations: we take $K=0.1$. We take $U=0.1$ as the elastic constant for the term constraining the length of the polarisation field---we have explored smaller and larger values of $U$, and find that sensible choices do not effect the equilibrium textures. The parameters $C, \lambda$ for the twist-bend nematic and $q_0$ for the cholesteric are chosen as described in the main text.

The simulated director and polarisation fields are output as VTK files and loaded into the software Paraview~\cite{hansen_visualization_2005}, which is used for all subsequent analysis and visualisation. 

Our main method for visualising the structures we observe is the Pontryagin--Thom (PT) construction~\cite{chen_topological_2012}. For this construction, we take the set of points where one component of the director field (we always use the $z$-component) vanishes. Generically this will be a surface. We then colour this surface according to the angle $\alpha$ between the other two components of the director. 

As with cholesterics, twist-bend nematics have a `pseudolayer structure' which can also help to visualise the structure. In cholesterics, the layers are surfaces to which the director is tangent or close to being tangent. In twist-bend nematics, the layers are the surfaces to which the bend is tangent. We can examine cholesteric layers near to a surface by plotting the dot product between the cholesteric director and the surface normal. In particular, the layer structure is well represented by the `dividing curve' on a surface that cuts across them~\cite{pollard_morse_2024}. This dividing curve is the set of points where the director is tangent to the surface. It captures the same topology as a PT surface, but unlike the PT surface it can be observed directly using light microscopy and hence it can be more helpful with comparing simulation with experiment. Moreover, by colouring a surface with normal ${\bf N}$ according to the dot product ${\bf n} \cdot {\bf N}$ we see both the layer structure and the Skyrmion charge across that surface, via Eq.~\eqref{eq:skyrmion}. The layer structure can also be seen by examining the magnitude of the bend vector field, which (as we show in several figures in the main text) reveals the same structure as the dividing curve. 

Defect points, where the director vanishes, necessarily appear on the PT surface as singularities in the angle $\alpha$ around which the colour `brushes' wind. We locate defects by finding these point on the PT surfaces, and represent them by coloured spheres. We determine the Morse index of each defect point by enclosing it in a spherical surface with normal ${\bf N}$ and plotting ${\bf n} \cdot {\bf N}$ to determine the dividing curve, as explained in Ref.~\cite{pollard_morse_2024}. 

The $\beta$-lines are located and visualised by plotting contours of $|{\bf b}|$~\cite{binysh_geometry_2020}. The value we use for the contour is selected for each texture based on what gives the clearest picture of the $\beta$-lines---typical values are $|{\bf b}| = 0.01-0.05$. In the cholesteric plots, $\lambda$-lines are similarly located using contours of $|\Delta|$, where $\Delta$ is the traceless and symmetric part the director gradients~\cite{efrati_orientation-dependent_2014, beller_geometry_2014, machon_umbilic_2016, selinger_interpretation_2018, pollard_intrinsic_2021, da_silva_moving_2021}. 

We also plot streamlines of the director field, which are calculated using Paraview's inbuilt `Stream Tracer' function. Arrows indicating the director and bend vector fields use the inbuilt `Glyph' function. 

\bibliography{references.bib}

\end{document}